\documentstyle{article}
\def\Journal#1#2#3#4{{#1} {\bf #2}, #3 (#4)}

\def\NPB{{\em Nucl. Phys.} B}

\def\PRL{\em Phys. Rev. Lett.}
\def\PRD{{\em Phys. Rev.} D}

\def\half{\mbox{$\frac{1}{2}$}}
\def\third{\mbox{$\frac{1}{3}$}}

\newcommand{\be}{\begin{equation}}
\newcommand{\ee}{\end{equation}}
\newcommand{\eq}[1]{Eq.~(\ref{#1})}
\newcommand{\order}{{\cal O}}
\newcommand{\nl}{\nonumber \\}
\newcommand{\bearray}{\begin{eqnarray}}
\newcommand{\eearray}{\end{eqnarray}}

\newcommand{\Tr}{{\rm Tr}}
\newcommand{\sw}{{\rm sw}}
\newcommand{\pv}{{\bf p}}

\title{Perturbative Improvement for Lattice QCD:\\
An Update\footnote{Invited talk at the workshop on {\em Lattice QCD on
Parallel Computers} (Tsukuba, March 1997).}}
\author{Peter Lepage\\
\small Newman Laboratory of Nuclear Studies, Cornell University\\
\small Ithaca, NY, 14853; email: gpl@mail.lns.cornell.edu}

\date{\small 18 July 1997}
\begin{document}

\maketitle

\begin{abstract}
Recent developments in the Symanzik improvement program for
lattice QCD are reviewed. 
%
\end{abstract}

\section{Introduction}
In this lecture I review recent trends in the effort to develop highly
improved discretizations of QCD for use on coarse lattices. I focus here on
the perturbative Symanzik approach to this problem; other 
approaches are discussed by other lecturers at this meeting. The ultimate
goal of all this activity is too allow us to simulate with larger lattice
spacings, at greatly reduced cost. The savings that
result from larger lattice spacings can be used to extend the reach of
lattice QCD to far more difficult problems, well suited to our largest
supercomputers. At the same time,
there is a wealth of simpler simulation problems that are now accessible to
anyone with a modern workstation or personal computer.

It is somewhat artificial to isolate a single approach like Symanzik
improvement. Symanzik improvement is but one of a large set of
tools\,---\,Symanzik and on-shell improvement, perturbation theory and
tadpole improvement, blocked fields and perfect actions, nonperturbative
tuning and MCRG, and so on\,---\,that we can use to design improved
discretizations. These tools are closely related, and not mutually exclusive.
Solving QCD is much too difficult for us to ignore
any particular set of tools; we need every trick available. 

I begin, in Section~\ref{symanzik-and-TI} with a review of perturbative
improvement and tadpole improvement. I discuss variations on the
canonical approach, including nonperturbative tuning, and I discuss what we
now know about the size of finite-$a$ errors. 
Weak-coupling perturbation theory plays a central role in all 
practical approaches to improved discretization. I discuss the status of
lattice perturbation theory, as well as how to implement and test it, in
Section~\ref{perturbation-theory}. In
Section~\ref{recent-developments} I discuss specific, recent
developments concerning various actions. Finally I
summarize in Section~\ref{conclusions}.

\section{Perturbative and Tadpole Improvement}\label{symanzik-and-TI}

\subsection{The Method}
The procedure for perturbatively improving a lattice QCD operator or
action involves three steps\,\cite{schladming}:
\begin{enumerate}
\item Classical Improvement:  The classical (tree-level) lattice
action or operator is corrected through the desired order in
the lattice spacing~$a$. For example, the lattice gluon action 
\bearray \label{gluon-action}
S &=& -\beta\sum_{x,\mu\!>\!\nu}\left\{ c_1 P_{\mu,\nu}(x) \right.\nl
&&+ c_2
\left.\left(R_{\mu\mu,\nu}(x) + R_{\nu\nu,\mu}(x)\right)\right\} 
\eearray
equals 
\be
 \int\! d^4x\,\half{\rm Tr}F_{\mu\nu}^2(x) + \order(a^4)
\ee
for classical fields when
\be
\beta = 6/g^2, \quad c_1 = 5/3, \quad c_2 = -1/12.
\ee
Here $P_{\mu,\nu}$ is one third the real part of the trace of the
$1\times1$~Wilson loop in the $\mu\nu$~plane, while $R_{\mu\mu,\nu}$
is the same for a $2\times1$ loop with two steps in the
$\mu$~direction and one in the $\nu$~direction. We derive this result
by substituting the definition of the lattice link, in terms
of the continuum vector potential,
\be
U_\mu(x) = {\rm P}\,{\rm exp}\left[-i\int_0^a\!\!d\xi\,
gA_\mu(x+\xi\hat\mu)\right],
\ee
into the lattice action, and expanding in powers of the lattice
spacing~$a$.

\item Tadpole Improvement: In the quantum theory, the couplings are
renormalized by quantum fluctuations, and new operators are required. The
dominant renormalizations are due to tadpole diagrams caused by the
(nonlinear) gluon link operator~$U_\mu$. These lattice artifacts are largely
cancelled by tadpole improvement: every link in the classical lattice theory
is divided by the mean link\,\cite{lm,schladming}:
\be
U_\mu(x) \to \frac{U_\mu(x)}{u_0}.
\ee
The mean link~$u_0$ is a number between 0~and~1; it is computed
self-consistently, using a Monte Carlo simulation, from the mean value of
$\third\Tr U_\mu$. This last quantity is gauge dependent and can be made
arbitrarily small by changing the gauge. Thus we choose the gauge that
maximizes $\langle\Tr U_\mu\rangle$, which is Landau gauge:
\be
u_0 \equiv \langle0| \third\Tr U_\mu|0 \rangle_{\rm LG}.
\ee
This gauge minimizes the tadpoles in $u_0$, pushing it as close to~1 as
possible; any tadpole contribution that remains cannot be a gauge artifact.
Our gluon action, \eq{gluon-action}, is tadpole improved by setting
\be
c_1 = \frac{5}{3\,u_0^4}, \quad\quad c_2 = -\,\frac{1}{12\,u_0^6}.
\ee
The tadpole correction here is substantial at lattice spacings of .25--.4\,fm  
since $u_0$ is .7--.8 and is raised to large powers.

At small lattice spacings $u_0$ is well approximated by the fourth
root~$u_p$ of the plaquette operator, which is easier to compute than
the mean link in Landau gauge:
\be \label{up}
u_0\approx u_p \equiv \langle0| P_{\mu,\nu} |0\rangle^{1/4}.
\ee

\item Radiative Corrections: 
A tadpole-improved classical action or operator is an efficient starting
point for a systematic weak-coupling expansion. Each coupling constant
has an expansion in powers of the QCD coupling
constant~$\alpha_V$:
\be
c_i = \frac{1}{u_0^{n_i}}\,( c_{i0} + c_{i1}\,\alpha_V(q^*_i) + \cdots)
\ee
The scale $q^*_i$ is typically between 0.5/a
and 2/a, and is easily computed using one-loop perturbation
theory\,\cite{lm}. The value of the coupling constant is also easily measured
in the simulation\,\cite{lm}. New operators, beyond those necessary in the
classical analysis, are usually needed when radiative corrections are
included. The perturbative expansions of the coupling constants are computed
by matching: low-energy physical quantities, scattering amplitudes for
example, are computed perturbatively in both the lattice theory and the
continuum, and lattice couplings are tuned so that the two calculations agree
order-by-order in the coupling (to a given order in the lattice spacing). The
perturbation expansion is usually much more convergent with the tadpole
factor, $1/u_0^{n_i}$, removed as above. Indeed, for many applications, only
the tree-level term in the expansion is necessary after tadpole improvement.
\end{enumerate}

The procedure just outlined
is a double expansion in powers of the lattice spacing and of
the QCD~coupling constant. Carried to all orders it becomes infinitely
complicated; in practice, however, only a few low orders of correction are
needed to reduce errors to a few percent. 
Recently we have started to obtain high-quality 
data on the size of finite-$a$ errors. 
An example is the SCRI analysis of errors in the hadron spectrum
for Wilson and SW quarks\,\cite{scri}. By going to large~$a$, they made the
errors large and unambiguous; they
tested the algorithms ``to failure.'' They find that the relative errors are
roughly $(a/a_0)^n$ with $n\!=\!1$ for Wilson quarks and $n\!=\!2$ for SW
quarks, and $a_0\approx$1--2\,fm. The powers are as expected, and so is the
scale~$a_0$ for the errors\,\cite{melbourne}. 

Such information tells us how much improvement we need at a given lattice
spacing. At $a\!=\!.4$\,fm, for example, $a/a_0\approx.4$ and
$\alpha_V\approx.3$.  Thus, if we want errors smaller than 5--10\%, we
should include in our quark action tree-level corrections through order~$a^3$
and radiative corrections through order~$a\,\alpha_V$ (that is, radiative
corrections for only the order~$a$ term, and then
only through one-loop). At $a\!=\!.1$\,fm, on the other hand, only the
order~$a$ and, perhaps, the order~$a\,\alpha_V$ corrections are necessary
for 1\%~accuracy (in addition to $4^6$ times as much computing). Note
that as 
$a$~increases higher powers of $a$ become more important faster
than higher powers of~$\alpha_V$. These
arguments are approximate, but they provide a starting point for a
realistic assessment of what is needed.

Radiative corrections (beyond tadpole improvement) are unimportant for most
terms in quark and gluon actions. This is because
the radiative corrections are in terms already suppressed by some
power of the lattice spacing; radiative corrections for terms of order~$a^0$
are generally absorbed into masses or other parameters that are tuned
numerically. The situation is different, however, for lattice
operators like currents. Typically 
the desired continuum current is obtained from a lattice current using
\be
J_{\rm cont} = Z(\alpha_V)\,\,J_{\rm lat} + 
\gamma_1(\alpha_V)\,a\,\delta J_1
+ \order(a^2).
\ee
While radiative corrections may be unimportant in the $\order(a,a^2\ldots)$
corrections, the $Z$-factor appears in order~$a^0$ and therefore
must usually be determined through one or two orders in~$\alpha_V$ to
achieve useful accuracy. The problem is even more accute for operators that
mix through radiative corrections with operators of lower dimension, because
of $1/a$ divergences; this can be alleviated to some extent by
working with larger~$a$'s, where $1/a$ is then smaller.

An alternative to tadpole improvement worth investigating 
is to replace the link operators in
quark actions and currents by smeared links: for example,
\be 
U_\mu \to \tilde{U}_\mu \equiv
\left[1+\frac{\epsilon\,a^2 \Delta^{(2)}}{n}\right]^n\,U_\mu
\ee
where $\Delta^{(2)}$ is the discretized
gauge-covariant laplacian, 
$n$~is a small integer, and $\epsilon$~a smearing parameter. This introduces
a factor like $\exp(-\epsilon\, a^2 q^2)$ into the gluon propagator which
reduces the effective ultraviolet cutoff for gluons coupled to the quarks.
Since tadpoles are quadratically divergent, even a modest reduction in this
cutoff greatly reduces their contribution and makes tadpole improvement
unnecessary. Substituting $\tilde{U}_\mu$ for $U_\mu$ in a lattice operator
introduces a new $a^2$~error; this can be removed by
including an additional factor of $(1-\epsilon\,a^2 \Delta^{(2)})$ 
in the definition of~$\tilde{U}_\mu$.

\subsection{Nonperturbative Tuning}
A variation on the standard procedure is to tune some of the couplings
nonperturbatively, using simulations. There are two standard approaches. One
is to tune a coupling constant to restore a
continuum symmetry. In many cases a leading correction term
breaks a continuum symmetry, like Lorentz invariance, and is the only
term of its order (in~$a$) to do so. Then the
coefficient of that term can be determined by adjusting it
until the corresponding symmetry is
restored in the simulation results. 
For example, there is only one independent dimension-six operator
that breaks rotation invariance and contributes to the
$\order(a^2)$~corrections in a gluon action. Thus we can adjust
coupling~$c_2$ in our lattice gluon action, \eq{gluon-action}, until, for
example, the simulation gives a static quark potential with
$V(2r,2r,r)=V(3r,0,0)$ for some~$r$. The tuned coupling then automatically
includes perturbative corrections to all orders in~$\alpha_V$, as well as any
nonperturbative contributions.\footnote{Any 
short-distance quantity could have nonperturbative contributions. 
Usually these are small compared 
with the perturbative contributions, which is why
perturbative QCD works so well. Note that lattice coupling constants have
renormalon contributions because of finite-$a$ errors in the lattice action;
but these contributions are automatically
suppressed to the same level as other finite-$a$ errors caused by the 
action, and
so can be ignored.} The tuned coupling is not exact, however, since the
quantities used to tune it have finite-$a$ errors caused by the 
action. In our example, the gluon action is accurate only
through order~$a^2$. Thus $V(2r,2r,r)-V(3r,0,0)$ has order $a^4$~errors;
tuning it to zero induces order~$a^4$~errors in $c_2$. Obviously, such
truncation errors are smaller for more highly
corrected lattice theories. Also some care is necessary in choosing
the quantitiy to be tuned; one wants a quantity for which finite-$a$ 
errors are minimal or small.

Other examples of couplings that can be tuned using symmetry restoration
include: the clover coefficient in the Sheikholeslami-Wohlert (SW) quark
action, using chiral symmetry\,\cite{desy-csw}; the coefficient of the
$\overline\psi \Delta^{(3)}\cdot\gamma\psi$ correction in the D234
actions, using
Lorentz invariance (see below); and $\kappa_t/\kappa_s$ in the Fermilab quark
action for massive quarks, using Lorentz invariance.

The second approach to nonperturbative tuning is to adjust couplings so
that physics remains the same at different lattice spacings. Thus, for
example, one might adjust the clover coefficient  in the SW action so that
the rho mass scales correctly from lattice spacing to lattice spacing. This
is  harder than tuning for a symmetry since simulations at several
lattice spacings are required, but it is not impossibly difficult if only
one or two couplings are needed. One loses one ``prediction'' from
the simulation for each coupling that is tuned this way. Again the couplings
that result include all radiative corrections and any nonperturbative
effects; and, again, the couplings have finite-$a$ errors caused by the
action. 

Nonperturbative tuning is particularly useful for leading-order corrections,
like the order-$a$ term in quark actions, where there are only one or two
couplings to tune. Where it can be used, it is also useful for
currents since otherwise one-loop and sometimes two-loop
radiative corrections must be computed. For finite-$a$ corrections beyond
leading order, however, tadpole improvement and perturbation theory are
likely to be the most feasible approach. Fortunately, tree-level perturbation
theory is probably adequate for most correction terms in order~$a^2$ and
higher.

\section{Perturbation Theory}\label{perturbation-theory}
\subsection{Does It Work?}

Weak-coupling perturbation theory plays an essential role in all
current approaches to improved lattice QCD actions and operators. 
It tells us that improving the classical theory is a good
starting point for the quantum improvement program. It tells us which
corrections are most important; as a
result we need only a small number of correction terms.  In the
perturbative improvement schemes I discuss here, perturbation theory
is the tool for calculating the coupling constants. But can we
trust perturbation theory?

\begin{figure}
\begin{center}
\setlength{\unitlength}{0.240900pt}
\ifx\plotpoint\undefined\newsavebox{\plotpoint}\fi
\sbox{\plotpoint}{\rule[-0.200pt]{0.400pt}{0.400pt}}%
\begin{picture}(900,475)(0,125)
\font\gnuplot=cmr10 at 10pt
\gnuplot
\sbox{\plotpoint}{\rule[-0.200pt]{0.400pt}{0.400pt}}%
\put(220.0,216.0){\rule[-0.200pt]{4.818pt}{0.400pt}}
\put(198,216){\makebox(0,0)[r]{$0.14$}}
\put(816.0,216.0){\rule[-0.200pt]{4.818pt}{0.400pt}}
\put(220.0,319.0){\rule[-0.200pt]{4.818pt}{0.400pt}}
\put(198,319){\makebox(0,0)[r]{$0.16$}}
\put(816.0,319.0){\rule[-0.200pt]{4.818pt}{0.400pt}}
\put(220.0,422.0){\rule[-0.200pt]{4.818pt}{0.400pt}}
\put(198,422){\makebox(0,0)[r]{$0.18$}}
\put(816.0,422.0){\rule[-0.200pt]{4.818pt}{0.400pt}}
\put(220.0,525.0){\rule[-0.200pt]{4.818pt}{0.400pt}}
\put(198,525){\makebox(0,0)[r]{$0.2$}}
\put(816.0,525.0){\rule[-0.200pt]{4.818pt}{0.400pt}}
\put(308.0,113.0){\rule[-0.200pt]{0.400pt}{4.818pt}}
\put(308,68){\makebox(0,0){$5$}}
\put(308.0,557.0){\rule[-0.200pt]{0.400pt}{4.818pt}}
\put(528.0,113.0){\rule[-0.200pt]{0.400pt}{4.818pt}}
\put(528,68){\makebox(0,0){$10$}}
\put(528.0,557.0){\rule[-0.200pt]{0.400pt}{4.818pt}}
\put(748.0,113.0){\rule[-0.200pt]{0.400pt}{4.818pt}}
\put(748,68){\makebox(0,0){$15$}}
\put(748.0,557.0){\rule[-0.200pt]{0.400pt}{4.818pt}}
\put(220.0,113.0){\rule[-0.200pt]{148.394pt}{0.400pt}}
\put(836.0,113.0){\rule[-0.200pt]{0.400pt}{111.778pt}}
\put(220.0,577.0){\rule[-0.200pt]{148.394pt}{0.400pt}}
\put(45,345){\makebox(0,0){$\alpha_{P}(q_{1,1})$}}
\put(528,23){\makebox(0,0){$q_{1,1}$ (GeV)}}
\put(220.0,113.0){\rule[-0.200pt]{0.400pt}{111.778pt}}
\put(264,527){\usebox{\plotpoint}}
\multiput(264.58,523.01)(0.497,-1.084){49}{\rule{0.120pt}{0.962pt}}
\multiput(263.17,525.00)(26.000,-54.004){2}{\rule{0.400pt}{0.481pt}}
\multiput(290.58,467.88)(0.497,-0.818){51}{\rule{0.120pt}{0.752pt}}
\multiput(289.17,469.44)(27.000,-42.439){2}{\rule{0.400pt}{0.376pt}}
\multiput(317.58,424.29)(0.497,-0.693){49}{\rule{0.120pt}{0.654pt}}
\multiput(316.17,425.64)(26.000,-34.643){2}{\rule{0.400pt}{0.327pt}}
\multiput(343.58,388.74)(0.497,-0.555){51}{\rule{0.120pt}{0.544pt}}
\multiput(342.17,389.87)(27.000,-28.870){2}{\rule{0.400pt}{0.272pt}}
\multiput(370.00,359.92)(0.519,-0.497){47}{\rule{0.516pt}{0.120pt}}
\multiput(370.00,360.17)(24.929,-25.000){2}{\rule{0.258pt}{0.400pt}}
\multiput(396.00,334.92)(0.591,-0.496){41}{\rule{0.573pt}{0.120pt}}
\multiput(396.00,335.17)(24.811,-22.000){2}{\rule{0.286pt}{0.400pt}}
\multiput(422.00,312.92)(0.713,-0.495){35}{\rule{0.668pt}{0.119pt}}
\multiput(422.00,313.17)(25.613,-19.000){2}{\rule{0.334pt}{0.400pt}}
\multiput(449.00,293.92)(0.768,-0.495){31}{\rule{0.712pt}{0.119pt}}
\multiput(449.00,294.17)(24.523,-17.000){2}{\rule{0.356pt}{0.400pt}}
\multiput(475.00,276.92)(0.849,-0.494){29}{\rule{0.775pt}{0.119pt}}
\multiput(475.00,277.17)(25.391,-16.000){2}{\rule{0.388pt}{0.400pt}}
\multiput(502.00,260.92)(1.012,-0.493){23}{\rule{0.900pt}{0.119pt}}
\multiput(502.00,261.17)(24.132,-13.000){2}{\rule{0.450pt}{0.400pt}}
\multiput(528.00,247.92)(1.012,-0.493){23}{\rule{0.900pt}{0.119pt}}
\multiput(528.00,248.17)(24.132,-13.000){2}{\rule{0.450pt}{0.400pt}}
\multiput(554.00,234.92)(1.251,-0.492){19}{\rule{1.082pt}{0.118pt}}
\multiput(554.00,235.17)(24.755,-11.000){2}{\rule{0.541pt}{0.400pt}}
\multiput(581.00,223.92)(1.203,-0.492){19}{\rule{1.045pt}{0.118pt}}
\multiput(581.00,224.17)(23.830,-11.000){2}{\rule{0.523pt}{0.400pt}}
\multiput(607.00,212.93)(1.543,-0.489){15}{\rule{1.300pt}{0.118pt}}
\multiput(607.00,213.17)(24.302,-9.000){2}{\rule{0.650pt}{0.400pt}}
\multiput(634.00,203.93)(1.485,-0.489){15}{\rule{1.256pt}{0.118pt}}
\multiput(634.00,204.17)(23.394,-9.000){2}{\rule{0.628pt}{0.400pt}}
\multiput(660.00,194.93)(1.682,-0.488){13}{\rule{1.400pt}{0.117pt}}
\multiput(660.00,195.17)(23.094,-8.000){2}{\rule{0.700pt}{0.400pt}}
\multiput(686.00,186.93)(1.748,-0.488){13}{\rule{1.450pt}{0.117pt}}
\multiput(686.00,187.17)(23.990,-8.000){2}{\rule{0.725pt}{0.400pt}}
\multiput(713.00,178.93)(1.942,-0.485){11}{\rule{1.586pt}{0.117pt}}
\multiput(713.00,179.17)(22.709,-7.000){2}{\rule{0.793pt}{0.400pt}}
\multiput(739.00,171.93)(2.018,-0.485){11}{\rule{1.643pt}{0.117pt}}
\multiput(739.00,172.17)(23.590,-7.000){2}{\rule{0.821pt}{0.400pt}}
\multiput(766.00,164.93)(1.942,-0.485){11}{\rule{1.586pt}{0.117pt}}
\multiput(766.00,165.17)(22.709,-7.000){2}{\rule{0.793pt}{0.400pt}}
\put(299,437){\circle{24}}
\put(475,277){\circle{24}}
\put(615,215){\circle{24}}
\put(715,166){\circle{24}}
\put(299.0,423.0){\rule[-0.200pt]{0.400pt}{6.745pt}}
\put(289.0,423.0){\rule[-0.200pt]{4.818pt}{0.400pt}}
\put(289.0,451.0){\rule[-0.200pt]{4.818pt}{0.400pt}}
\put(475.0,273.0){\rule[-0.200pt]{0.400pt}{2.168pt}}
\put(465.0,273.0){\rule[-0.200pt]{4.818pt}{0.400pt}}
\put(465.0,282.0){\rule[-0.200pt]{4.818pt}{0.400pt}}
\put(615.0,207.0){\rule[-0.200pt]{0.400pt}{3.854pt}}
\put(605.0,207.0){\rule[-0.200pt]{4.818pt}{0.400pt}}
\put(605.0,223.0){\rule[-0.200pt]{4.818pt}{0.400pt}}
\put(715.0,140.0){\rule[-0.200pt]{0.400pt}{12.286pt}}
\put(705.0,140.0){\rule[-0.200pt]{4.818pt}{0.400pt}}
\put(705.0,191.0){\rule[-0.200pt]{4.818pt}{0.400pt}}
\end{picture}
\end{center}
\caption{The running coupling constant~$\alpha_P(q)$,
from the plaquette at different lattice spacings, plotted versus the
typical momentum, $q_{1,1} = 3.4/a$, in the plaquette; lattice
spacings are from the 1$S$--1$P$ upsilon splittin in NRQCD. The solid
line is predicted by the three-loop beta function.}
\label{fig:alpha-evol}
\end{figure}
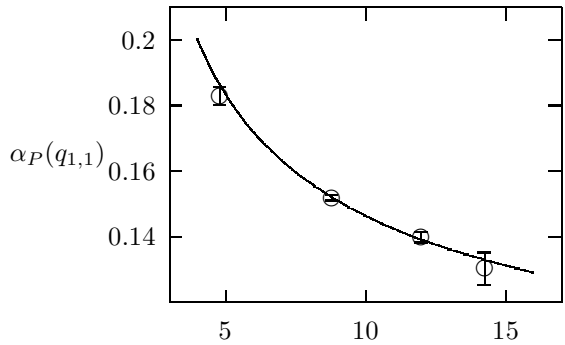

The short answer is yes. Wherever careful tests have been made, using
perturbative expansions through second or third order, for quantities that
are easy to measure accurately in simulations, perturbation theory has been
successful. For example, in~\cite{NRQCD-alpha} four values for the QCD coupling
constant~$\alpha_P$
are extracted using third-order expansions for four different
Wilson loops\,---\,$W_{11}$, $W_{12}$, $W_{13}$, and $W_{22}$\,---\,at
lattice spacing $a\!\approx\!0.1$\,fm ($\beta\!=\!6$). These agree with
each other to within the estimated truncation errors,
$O(\alpha_P^4) \approx 1\%$; and they agree to the same
precision with the value obtained from the (very different) Schr\"odinger
Functional approach\,\cite{weisz}. This is a high-precision
test of QCD perturbation theory, much more stringent than any test based
on continuum quantities.  
Another three-loop test is coupling constant evolution: As illustrated in
Fig~\ref{fig:alpha-evol}, the running of the QCD coupling constant
is well described by  the perturbative beta function
at least down to scales of order 5\,GeV;
less accurate tests show this is true even
down to scales of order~1\,GeV\,\cite{lm}.

\begin{figure}
\begin{center}
\setlength{\unitlength}{0.240900pt}
\ifx\plotpoint\undefined\newsavebox{\plotpoint}\fi
\sbox{\plotpoint}{\rule[-0.200pt]{0.400pt}{0.400pt}}%
\begin{picture}(900,475)(0,125)
\font\gnuplot=cmr10 at 10pt
\gnuplot
\sbox{\plotpoint}{\rule[-0.200pt]{0.400pt}{0.400pt}}%
\put(220.0,206.0){\rule[-0.200pt]{4.818pt}{0.400pt}}
\put(198,206){\makebox(0,0)[r]{$0.13$}}
\put(816.0,206.0){\rule[-0.200pt]{4.818pt}{0.400pt}}
\put(220.0,299.0){\rule[-0.200pt]{4.818pt}{0.400pt}}
\put(198,299){\makebox(0,0)[r]{$0.135$}}
\put(816.0,299.0){\rule[-0.200pt]{4.818pt}{0.400pt}}
\put(220.0,391.0){\rule[-0.200pt]{4.818pt}{0.400pt}}
\put(198,391){\makebox(0,0)[r]{$0.14$}}
\put(816.0,391.0){\rule[-0.200pt]{4.818pt}{0.400pt}}
\put(220.0,484.0){\rule[-0.200pt]{4.818pt}{0.400pt}}
\put(198,484){\makebox(0,0)[r]{$0.145$}}
\put(816.0,484.0){\rule[-0.200pt]{4.818pt}{0.400pt}}
\put(276.0,113.0){\rule[-0.200pt]{0.400pt}{4.818pt}}
\put(276,68){\makebox(0,0){$6$}}
\put(276.0,557.0){\rule[-0.200pt]{0.400pt}{4.818pt}}
\put(388.0,113.0){\rule[-0.200pt]{0.400pt}{4.818pt}}
\put(388,68){\makebox(0,0){$6.2$}}
\put(388.0,557.0){\rule[-0.200pt]{0.400pt}{4.818pt}}
\put(500.0,113.0){\rule[-0.200pt]{0.400pt}{4.818pt}}
\put(500,68){\makebox(0,0){$6.4$}}
\put(500.0,557.0){\rule[-0.200pt]{0.400pt}{4.818pt}}
\put(612.0,113.0){\rule[-0.200pt]{0.400pt}{4.818pt}}
\put(612,68){\makebox(0,0){$6.6$}}
\put(612.0,557.0){\rule[-0.200pt]{0.400pt}{4.818pt}}
\put(724.0,113.0){\rule[-0.200pt]{0.400pt}{4.818pt}}
\put(724,68){\makebox(0,0){$6.8$}}
\put(724.0,557.0){\rule[-0.200pt]{0.400pt}{4.818pt}}
\put(220.0,113.0){\rule[-0.200pt]{148.394pt}{0.400pt}}
\put(836.0,113.0){\rule[-0.200pt]{0.400pt}{111.778pt}}
\put(220.0,577.0){\rule[-0.200pt]{148.394pt}{0.400pt}}
\put(45,345){\makebox(0,0){$\kappa_c$}}
\put(528,23){\makebox(0,0){$\beta$}}
\put(220.0,113.0){\rule[-0.200pt]{0.400pt}{111.778pt}}
\put(706,512){\makebox(0,0)[r]{MC}}
\put(750,512){\circle{24}}
\put(276,302){\circle{24}}
\put(388,313){\circle{24}}
\put(500,313){\circle{24}}
\put(724,300){\circle{24}}
\put(706,467){\makebox(0,0)[r]{P.Th.}}
\put(750,467){\circle*{18}}
\put(276,362){\circle*{18}}
\put(388,347){\circle*{18}}
\put(500,334){\circle*{18}}
\put(724,312){\circle*{18}}
\put(728.0,467.0){\rule[-0.200pt]{15.899pt}{0.400pt}}
\put(728.0,457.0){\rule[-0.200pt]{0.400pt}{4.818pt}}
\put(794.0,457.0){\rule[-0.200pt]{0.400pt}{4.818pt}}
\put(276.0,310.0){\rule[-0.200pt]{0.400pt}{25.054pt}}
\put(266.0,310.0){\rule[-0.200pt]{4.818pt}{0.400pt}}
\put(266.0,414.0){\rule[-0.200pt]{4.818pt}{0.400pt}}
\put(388.0,304.0){\rule[-0.200pt]{0.400pt}{20.717pt}}
\put(378.0,304.0){\rule[-0.200pt]{4.818pt}{0.400pt}}
\put(378.0,390.0){\rule[-0.200pt]{4.818pt}{0.400pt}}
\put(500.0,297.0){\rule[-0.200pt]{0.400pt}{17.827pt}}
\put(490.0,297.0){\rule[-0.200pt]{4.818pt}{0.400pt}}
\put(490.0,371.0){\rule[-0.200pt]{4.818pt}{0.400pt}}
\put(724.0,282.0){\rule[-0.200pt]{0.400pt}{14.213pt}}
\put(714.0,282.0){\rule[-0.200pt]{4.818pt}{0.400pt}}
\put(714.0,341.0){\rule[-0.200pt]{4.818pt}{0.400pt}}
\end{picture}
\end{center}
\caption{Critical $\kappa$ for SW quarks from Monte Carlo simulations and
from first-order perturbation theory. Statistical errors are negligible for
the Monte Carlo results; relative errors shown for the perturbation theory
are $\pm1\alpha_V^2$.}
\label{fig:kappac}
\end{figure}
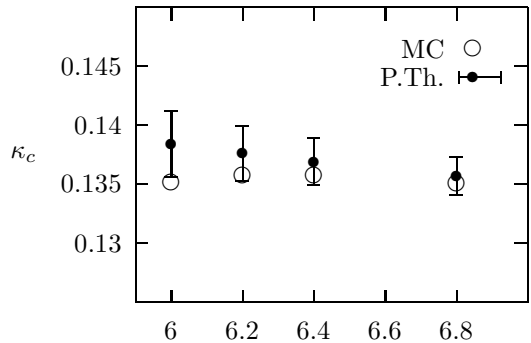

Usually when we use perturbation theory, we truncate the expansion at some
finite order in~$\alpha_V$. We should always estimate the truncation error
that results. Sometimes an apparent 
disagreement between perturbation theory and simulation results
is negligible when compared with the truncation error, and is therefore of
little significance. For example, the critical~$\kappa$ for SW quarks in
tadpole-improved perturbation theory is known only through
first-order\,\cite{desy-csw,colin-kappa}:
\be
\kappa_c = \frac{1}{8u_p}\left[1-0.20\,\alpha_V(4.4/a) +
\order(\alpha_V^2)\right]
\ee
where $u_p$ is the plaquette mean link. Thus perturbation theory should only
be correct up to relative errors of order~$\pm1\alpha_V^2$. 
In Figure~\ref{fig:kappac}, we compare
the first-order result, including error bars for higher-order terms,
with  nonperturbative results obtained from simulations\,\cite{desy-csw}.
The nonperturbative results are more accurate, but
perturbation theory agrees well\,---\,that is, it agrees
to within truncation errors. 

\begin{figure}
\begin{center}
\setlength{\unitlength}{0.240900pt}
\ifx\plotpoint\undefined\newsavebox{\plotpoint}\fi
\sbox{\plotpoint}{\rule[-0.200pt]{0.400pt}{0.400pt}}%
\begin{picture}(900,475)(0,125)
\font\gnuplot=cmr10 at 10pt
\gnuplot
\sbox{\plotpoint}{\rule[-0.200pt]{0.400pt}{0.400pt}}%
\put(220.0,190.0){\rule[-0.200pt]{4.818pt}{0.400pt}}
\put(198,190){\makebox(0,0)[r]{$1.2$}}
\put(816.0,190.0){\rule[-0.200pt]{4.818pt}{0.400pt}}
\put(220.0,268.0){\rule[-0.200pt]{4.818pt}{0.400pt}}
\put(198,268){\makebox(0,0)[r]{$1.4$}}
\put(816.0,268.0){\rule[-0.200pt]{4.818pt}{0.400pt}}
\put(220.0,345.0){\rule[-0.200pt]{4.818pt}{0.400pt}}
\put(198,345){\makebox(0,0)[r]{$1.6$}}
\put(816.0,345.0){\rule[-0.200pt]{4.818pt}{0.400pt}}
\put(220.0,422.0){\rule[-0.200pt]{4.818pt}{0.400pt}}
\put(198,422){\makebox(0,0)[r]{$1.8$}}
\put(816.0,422.0){\rule[-0.200pt]{4.818pt}{0.400pt}}
\put(220.0,500.0){\rule[-0.200pt]{4.818pt}{0.400pt}}
\put(198,500){\makebox(0,0)[r]{$2$}}
\put(816.0,500.0){\rule[-0.200pt]{4.818pt}{0.400pt}}
\put(308.0,113.0){\rule[-0.200pt]{0.400pt}{4.818pt}}
\put(308,68){\makebox(0,0){$5.8$}}
\put(308.0,557.0){\rule[-0.200pt]{0.400pt}{4.818pt}}
\put(396.0,113.0){\rule[-0.200pt]{0.400pt}{4.818pt}}
\put(396,68){\makebox(0,0){$6$}}
\put(396.0,557.0){\rule[-0.200pt]{0.400pt}{4.818pt}}
\put(484.0,113.0){\rule[-0.200pt]{0.400pt}{4.818pt}}
\put(484,68){\makebox(0,0){$6.2$}}
\put(484.0,557.0){\rule[-0.200pt]{0.400pt}{4.818pt}}
\put(572.0,113.0){\rule[-0.200pt]{0.400pt}{4.818pt}}
\put(572,68){\makebox(0,0){$6.4$}}
\put(572.0,557.0){\rule[-0.200pt]{0.400pt}{4.818pt}}
\put(660.0,113.0){\rule[-0.200pt]{0.400pt}{4.818pt}}
\put(660,68){\makebox(0,0){$6.6$}}
\put(660.0,557.0){\rule[-0.200pt]{0.400pt}{4.818pt}}
\put(748.0,113.0){\rule[-0.200pt]{0.400pt}{4.818pt}}
\put(748,68){\makebox(0,0){$6.8$}}
\put(748.0,557.0){\rule[-0.200pt]{0.400pt}{4.818pt}}
\put(220.0,113.0){\rule[-0.200pt]{148.394pt}{0.400pt}}
\put(836.0,113.0){\rule[-0.200pt]{0.400pt}{111.778pt}}
\put(220.0,577.0){\rule[-0.200pt]{148.394pt}{0.400pt}}
\put(45,345){\makebox(0,0){$c_{\rm sw}$}}
\put(528,23){\makebox(0,0){$\beta$}}
\put(220.0,113.0){\rule[-0.200pt]{0.400pt}{111.778pt}}
\put(706,512){\makebox(0,0)[r]{MC}}
\put(728.0,512.0){\rule[-0.200pt]{15.899pt}{0.400pt}}
\put(396,376){\usebox{\plotpoint}}
\multiput(396.00,374.92)(0.882,-0.498){97}{\rule{0.804pt}{0.120pt}}
\multiput(396.00,375.17)(86.331,-50.000){2}{\rule{0.402pt}{0.400pt}}
\multiput(484.00,324.92)(1.643,-0.497){51}{\rule{1.404pt}{0.120pt}}
\multiput(484.00,325.17)(85.087,-27.000){2}{\rule{0.702pt}{0.400pt}}
\multiput(572.00,297.92)(2.865,-0.497){59}{\rule{2.371pt}{0.120pt}}
\multiput(572.00,298.17)(171.079,-31.000){2}{\rule{1.185pt}{0.400pt}}
\put(396,446){\usebox{\plotpoint}}
\multiput(396.00,444.92)(0.595,-0.499){145}{\rule{0.576pt}{0.120pt}}
\multiput(396.00,445.17)(86.805,-74.000){2}{\rule{0.288pt}{0.400pt}}
\multiput(484.00,370.92)(1.133,-0.498){75}{\rule{1.003pt}{0.120pt}}
\multiput(484.00,371.17)(85.919,-39.000){2}{\rule{0.501pt}{0.400pt}}
\multiput(572.00,331.92)(1.923,-0.498){89}{\rule{1.630pt}{0.120pt}}
\multiput(572.00,332.17)(172.616,-46.000){2}{\rule{0.815pt}{0.400pt}}
\put(706,467){\makebox(0,0)[r]{PTh $u_p$}}
\put(750,467){\circle*{18}}
\put(396,326){\circle*{18}}
\put(484,308){\circle*{18}}
\put(572,294){\circle*{18}}
\put(748,272){\circle*{18}}
\put(728.0,467.0){\rule[-0.200pt]{15.899pt}{0.400pt}}
\put(728.0,457.0){\rule[-0.200pt]{0.400pt}{4.818pt}}
\put(794.0,457.0){\rule[-0.200pt]{0.400pt}{4.818pt}}
\put(396.0,291.0){\rule[-0.200pt]{0.400pt}{16.863pt}}
\put(386.0,291.0){\rule[-0.200pt]{4.818pt}{0.400pt}}
\put(386.0,361.0){\rule[-0.200pt]{4.818pt}{0.400pt}}
\put(484.0,281.0){\rule[-0.200pt]{0.400pt}{13.009pt}}
\put(474.0,281.0){\rule[-0.200pt]{4.818pt}{0.400pt}}
\put(474.0,335.0){\rule[-0.200pt]{4.818pt}{0.400pt}}
\put(572.0,274.0){\rule[-0.200pt]{0.400pt}{9.395pt}}
\put(562.0,274.0){\rule[-0.200pt]{4.818pt}{0.400pt}}
\put(562.0,313.0){\rule[-0.200pt]{4.818pt}{0.400pt}}
\put(748.0,256.0){\rule[-0.200pt]{0.400pt}{7.468pt}}
\put(738.0,256.0){\rule[-0.200pt]{4.818pt}{0.400pt}}
\put(738.0,287.0){\rule[-0.200pt]{4.818pt}{0.400pt}}
\put(706,422){\makebox(0,0)[r]{PTh $u_0$}}
\put(750,422){\circle{18}}
\put(396,401){\circle{18}}
\put(484,362){\circle{18}}
\put(572,337){\circle{18}}
\put(728.0,422.0){\rule[-0.200pt]{15.899pt}{0.400pt}}
\put(728.0,412.0){\rule[-0.200pt]{0.400pt}{4.818pt}}
\put(794.0,412.0){\rule[-0.200pt]{0.400pt}{4.818pt}}
\put(396.0,362.0){\rule[-0.200pt]{0.400pt}{18.790pt}}
\put(386.0,362.0){\rule[-0.200pt]{4.818pt}{0.400pt}}
\put(386.0,440.0){\rule[-0.200pt]{4.818pt}{0.400pt}}
\put(484.0,335.0){\rule[-0.200pt]{0.400pt}{13.009pt}}
\put(474.0,335.0){\rule[-0.200pt]{4.818pt}{0.400pt}}
\put(474.0,389.0){\rule[-0.200pt]{4.818pt}{0.400pt}}
\put(572.0,314.0){\rule[-0.200pt]{0.400pt}{11.081pt}}
\put(562.0,314.0){\rule[-0.200pt]{4.818pt}{0.400pt}}
\put(562.0,360.0){\rule[-0.200pt]{4.818pt}{0.400pt}}
\put(264,375){\circle*{18}}
\put(264.0,294.0){\rule[-0.200pt]{0.400pt}{39.026pt}}
\put(254.0,294.0){\rule[-0.200pt]{4.818pt}{0.400pt}}
\put(254.0,456.0){\rule[-0.200pt]{4.818pt}{0.400pt}}
\put(264,531){\circle{18}}
\put(264.0,430.0){\rule[-0.200pt]{0.400pt}{35.412pt}}
\put(254.0,430.0){\rule[-0.200pt]{4.818pt}{0.400pt}}
\put(254.0,577.0){\rule[-0.200pt]{4.818pt}{0.400pt}}
\end{picture}
\end{center}
\caption{Clover coefficient~$c_\sw$ for SW quarks from Monte Carlo
simulations, and from first-order perturbation theory using the plaquette
mean link~$u_p$ and the Landau-gauge mean link~$u_0$. Relative errors shown
for the perturbation theory are $\pm1\alpha_V^2$. The lines indicate
the possible range of the nonperturbative~$c_\sw$ (due to finite-$a$
errors) around the parameterization given in~\protect\cite{desy-csw}.}
\label{fig:csw}
\end{figure}
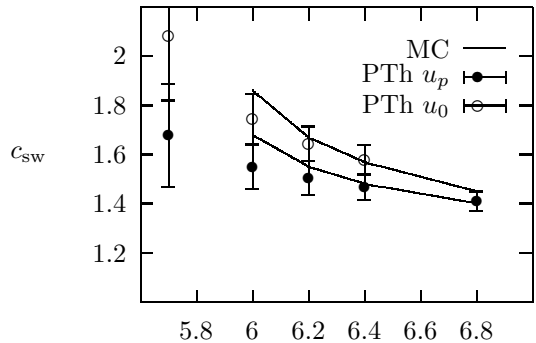

It is equally important, when using nonperturbative
tuning, to estimate the finite-$a$ errors due to truncation of the lattice
action. The clover coefficient~$c_\sw$ in the SW action can be tuned
nonperturbatively using chiral symmetry\,\cite{desy-csw}.
The tuned~$c_\sw$, however, has truncation errors of relative order $a/a_0$
for some $a_0$ around 1--2\,fm, at least for larger lattice spacings
(at smaller lattice spacings one can compute the order~$a$ term, using
perturbation theory, and
remove it). The tuned coupling can be compared with the perturbative
result\,\cite{desy-csw}, 
\be
c_{\rm sw} = \frac{1}{u_p^3} \left[ 1 + 0.20\,\alpha_V(q^*_\sw) 
+\order(\alpha_V^2)\right],
\ee
which also has truncation errors, of relative order~$\pm1\alpha_V^2$.
Scale~$q_\sw^*$ is currently unknown, but is likely around $.5/a$--$2/a$. In
Fig.~\ref{fig:csw} we compare the nonperturbative and perturbative results,
indicating the uncertainties in each, for a range of~$\beta$'s. (For this we
guess $a_0\!=\!2$\,fm and
$q_\sw^*\!=\!1/a$). The perturbative and nonperturbative
$c_\sw$'s agree well, and are roughly equal in accuracy; each
reinforces our confidence in the other. Note that, since
$c_\sw$ is the coefficient of an $\order(a)$~correction, 
the different $c_\sw$'s at $\beta\!=\!6$ give identical hadronic
physics to within~1--2\%.

We also include results in the last figure from first-order perturbation
theory, but with the Landau-gauge mean link~$u_0$ instead of~$u_p$:
\be
c_{\rm sw} = \frac{1}{u_0^3} \left[ 1 + 0.46\,\alpha_V 
+\order(\alpha_V^2)\right].
\label{csw-sw}
\ee
While $u_0$ and $u_p$ are almost identical through order~$\alpha_V$, they
differ in higher order. Thus differences between the two options for tadpole
improvement indicate the potential size of order~$\alpha_V^2$ corrections;
these differences are 5--10~times smaller if
$\alpha_V^2$~corrections are included in the two $c_\sw$~expansions. 

Perturbation theory triumphs in these and most other examples; it
is usually neither more nor less accurate than expected (from perturbation
theory). Nevertheless there will be situations where perturbation theory
fails to converge. In QED, for example, the decay rate for orthopositronium
is proportional to $\alpha^6\left(1-10.3\alpha/\pi\right)$\,\cite{cls}. This
expansion would be useless if, in QED, $\alpha=$~0.2 or~0.3. QCD
perturbation theory will fail at  distances smaller than 0.2\,fm, but not
often and not by much; one should worry about one's simulation if it
disagrees with
perturbation theory at such distances.

\subsection{Keeping Up}
Improved lattice actions and operators can be quite complicated, and lately
they have been changing rapidly. Unfortunately  perturbative calculations of
coupling constants and renormalization constants must be redone if any of
the actions or operators involved is changed. Also the vertices and
propagators can be quite complicated in a highly improved theory, making
even one-loop calculations a chore. The perturbative work on NRQCD, which
has a relatively 
complicated action, demonstrates that such calculations are
possible\,\cite{shigemitsu}. 
We need to systematize such calculations so that
one-loop calculations, at least, can be completed in a short time for any
new quark or gluon action. Otherwise we will find ourselves tied to older,
probably much less efficient actions and operators by the lack of
perturbative calculations for the new ones. Substantial progress in this
direction is being made now, but more effort is needed.

One approach that deals with this problem is to use Monte Carlo simulations
to do the lattice perturbation theory\,\cite{dimm}. On-shell 
quark form factors,
renormalization constants, masses, etc.\ can be computed using
simulations at very small lattice spacings, for example 
$a\!\approx\!0.002$\,fm at $\beta\!=\!9$, and matched directly with continuum
calculations. Since confinement is irrelevant at such short distances,
coupling constants determined in such a matching can be fit by polynomials
in~$\alpha_V$ (which is also 
measured in the simulation). In this way we obtain
perturbative expansions for the couplings that can be used in simulations
at more normal lattice spacings.

\section{Recent Developments}\label{recent-developments}

\subsection{Landau-Gauge Mean Link}
It is standard practice to approximate the Landau-gauge mean 
link~$u_0$ by the plaquette mean link~$u_p$. When the lattice
spacing is small, either of these is a reasonable choice. The two
definitions differ, however, by almost 10\% when $a\!=\!0.4$\,fm
(probably due mostly to perturbative differences in
$\order(\alpha_V^2)$). While coupling constants obtained from the
two definitions agree when 
perturbative corrections are included to all orders, tree-level or
one-loop results can be quite different when $a$~is large. Three
pieces of evidence suggest that it is better to stay with the Landau-gauge
mean link~$u_0$. First the tree-level,
tadpole-improved gluon action in~\eq{gluon-action} shows 2--3~times less
violation of rotation invariance (in the static potential) when $u_0$ is used
rather than~$u_p$\,\cite{gpl-u0}. Second, 
scaling violations in the charmonium groundstate hyperfine splitting are
3~times smaller when the tree-level
$\sigma\cdot B$ interaction is tadpole-improved with~$u_0$ rather
than~$u_p$\,\cite{trottier-v6}. 
Finally, the nonperturbative determination of the clover
coefficient~$c_\sw$ for SW quarks agrees more closely with
tadpole-improved one-loop
perturbation theory when $u_0$ is used, as is evident in
Fig.~\ref{fig:csw}. The same has also been observed at larger lattice
spacings, with improved gluon actions, and also in improved quark actions like
the D234c action discussed below\,\cite{d234c}.  These examples suggest that
$u_0$ might be the better choice generally, at least for large lattice
spacings, if only tree-level or first-order corrections are included;
there is little difference if
corrections through second order are included.

\subsection{Anisotropic Lattices}
It is widely felt that coarse lattices are noisier than fine
lattices. This is correct, especially for 
correlation functions where signal/noise vanishes expontially quickly
with increasing separation~$t$. It is true despite the fact that
the ratio of signal to noise is approximately
independent of lattice spacing when $t$~is large;
noise overcomes signal at some fixed physical distance, independent of
the lattice spacing. The problem on coarse lattices is that this fixed
distance might be only one or two lattice spacings long, since the
lattice spacing is large. Consequently, without large Monte Carlo data
sets, the statistically usable part of a mass
plateau, for example, may be too short to allow identification or
verification of the plateau. 
A lattice with small lattice spacing allows us to identify the plateau
at smaller distances, with less noise, but at increased cost from the
small lattice spacing. Since signal/noise decreases
exponentially with increasing~$t$, 
while simulation cost grows only as a
(large) power of lattice spacing, 
one might decide that smaller lattice spacings are
ultimately more efficient. Much of the advantage of the larger lattice
spacing is regained by careful design of sources and sinks, but a
far more efficient approach is to reduce
the lattice spacing~$a_t$ in the time direction while retaining a
large spacing~$a_s$ in spatial directions. Then more values of a 
correlation function can be measured at small~$t$'s, where relatively
few Monte Carlo measurements are needed to suppress the noise. The small 
increase in simulation cost that results
from a smaller~$a_t$ is more than offset by the
exponential improvement in signal-to-noise at the smaller~$t$'s.

Anisotropic actions are easy to design. The standard improved gluon
action becomes\,\cite{aniso,schladming}
\begin{eqnarray} 
{ S}_{n} &=& - \beta \sum_{x,\,s>s^\prime}
\frac{a_t}{a_s}\,\eta \,\left\{ 
\frac{5}{3}  \frac{P_{s,s^\prime}}{u_s^4} \right. \\
&& \quad\quad \left. - \frac{1}{12} \frac{R_{ss,s^\prime}}{u_s^6} 
- \frac{1}{12} \frac{R_{s^\prime s^\prime, s}}{u_s^6}
\right\} \\ \nonumber
&& - \beta \sum_{x,\,s}
\frac{a_s}{a_t} \,\eta^{-1}\,\left\{ 
\frac{4}{3}  \frac{P_{s,t}}{u_s^2 u_t^2}
- \frac{1}{12} \frac{R_{ss,t}}{u_s^4 u_t^2}
\right\},
\label{aniso-gluon-action}
\end{eqnarray}
where $u_s$ and $u_t$ are the (Landau-gauge) spatial and temporal mean
links, respectively, and $\eta$ accounts for renormalization of
the anisotropy by quantum effects. 
Since $a_t\!\ll\!a_s$, we have not bothered to
correct for order-$a_t^2$ errors in this action; thus the action
extends at most one step in the time direction and has no gluon
ghosts. An $a^2$-improved version of Landau gauge is obtained by
maximizing
\begin{equation}
\sum_{x\,\mu} \frac{1}{u_\mu a_\mu^2}\,{\rm ReTr}\left\{
U_\mu(x) - \frac{1}{16u_\mu}\,U_\mu(x)\,U_\mu(x+\hat\mu)\right\}.
\end{equation}

The anisotropic gluon 
action seems to have six parameters, but these appear in only 
three independent combinations,
\be
\frac{\beta}{u_s^3 u_t}, \quad\quad \left(\frac{\eta a_t}{a_s}\right)
\frac{u_t}{u_s}, \quad\quad u_s,
\ee
which control the spatial lattice spacing, the anisotropy, and
the size of the order~$a_s^2$ corrections, respectively. These can be
tuned numerically by choosing $\beta$ and $\eta a_t/a_s$, then
determining $u_t$ and $u_s$ self-consistently, and finally
measuring~$\eta$. Alternatively one might compute the last
three parameters using perturbation theory. In~\cite{aniso}
simulations with very small lattice spacing are combined with
first-order perturbation theory to obtain (prelimary) perturbative
expressions for the case $\eta a_t/a_s = 0.5$:
\bearray
u_s &=& 1 - 0.517/\beta + .17(5)/\beta^2 \\
u_t &=& 1 - 0.105/\beta - .02(1)/\beta^2 \\
\eta &=& 1-0.114(7)/\beta.
\eearray
These expressions give good results even at $a\!=\!.4$\,fm, 
as shown in Table~\ref{tab:anisoV}.
\begin{table}
\begin{center}
\begin{tabular}{llll}
\hline
&&$\beta\!=\!1.8$ & $\beta\!=\!1.34$ \\ \hline
$u_s$ & P.Th. & .77(2) & .71(3) \\
  & M.C. & .75 & .69 \\ \\
$u_t$ & P.Th. & .94(1) & .91(1) \\
  & M.C. & .93 & .91 \\ \\
$\eta$& P.Th. & .94(1) & .91(2) \\
  & M.C. & .94(1) & .90(1) \\
 \hline
\end{tabular}
\end{center}
\caption{Monte Carlo and perturbation theory values for parameters in
the anisotropic gluon action with $\eta a_t/a_s = 0.5$. Results are
given for two lattice spacings (approximately 0.25\,fm and .4\,fm),
and are preliminary.}
\label{tab:anisoV}
\end{table}

\begin{figure}
\begin{center}
\setlength{\unitlength}{0.240900pt}
\ifx\plotpoint\undefined\newsavebox{\plotpoint}\fi
\sbox{\plotpoint}{\rule[-0.200pt]{0.400pt}{0.400pt}}%
\begin{picture}(900,624)(0,125)
\font\gnuplot=cmr10 at 10pt
\gnuplot
\sbox{\plotpoint}{\rule[-0.200pt]{0.400pt}{0.400pt}}%
\put(220.0,113.0){\rule[-0.200pt]{148.394pt}{0.400pt}}
\put(220.0,113.0){\rule[-0.200pt]{0.400pt}{147.672pt}}
\put(220.0,266.0){\rule[-0.200pt]{4.818pt}{0.400pt}}
\put(198,266){\makebox(0,0)[r]{$1$}}
\put(816.0,266.0){\rule[-0.200pt]{4.818pt}{0.400pt}}
\put(220.0,420.0){\rule[-0.200pt]{4.818pt}{0.400pt}}
\put(198,420){\makebox(0,0)[r]{$2$}}
\put(816.0,420.0){\rule[-0.200pt]{4.818pt}{0.400pt}}
\put(220.0,573.0){\rule[-0.200pt]{4.818pt}{0.400pt}}
\put(198,573){\makebox(0,0)[r]{$3$}}
\put(816.0,573.0){\rule[-0.200pt]{4.818pt}{0.400pt}}
\put(308.0,113.0){\rule[-0.200pt]{0.400pt}{4.818pt}}
\put(308,68){\makebox(0,0){$1$}}
\put(308.0,706.0){\rule[-0.200pt]{0.400pt}{4.818pt}}
\put(396.0,113.0){\rule[-0.200pt]{0.400pt}{4.818pt}}
\put(396,68){\makebox(0,0){$2$}}
\put(396.0,706.0){\rule[-0.200pt]{0.400pt}{4.818pt}}
\put(484.0,113.0){\rule[-0.200pt]{0.400pt}{4.818pt}}
\put(484,68){\makebox(0,0){$3$}}
\put(484.0,706.0){\rule[-0.200pt]{0.400pt}{4.818pt}}
\put(572.0,113.0){\rule[-0.200pt]{0.400pt}{4.818pt}}
\put(572,68){\makebox(0,0){$4$}}
\put(572.0,706.0){\rule[-0.200pt]{0.400pt}{4.818pt}}
\put(660.0,113.0){\rule[-0.200pt]{0.400pt}{4.818pt}}
\put(660,68){\makebox(0,0){$5$}}
\put(660.0,706.0){\rule[-0.200pt]{0.400pt}{4.818pt}}
\put(748.0,113.0){\rule[-0.200pt]{0.400pt}{4.818pt}}
\put(748,68){\makebox(0,0){$6$}}
\put(748.0,706.0){\rule[-0.200pt]{0.400pt}{4.818pt}}
\put(220.0,113.0){\rule[-0.200pt]{148.394pt}{0.400pt}}
\put(836.0,113.0){\rule[-0.200pt]{0.400pt}{147.672pt}}
\put(220.0,726.0){\rule[-0.200pt]{148.394pt}{0.400pt}}
\put(89,419){\makebox(0,0){$a_s\,V$}}
\put(528,23){\makebox(0,0){$r/a_s$}}
\put(220.0,113.0){\rule[-0.200pt]{0.400pt}{147.672pt}}
\sbox{\plotpoint}{\rule[-0.500pt]{1.000pt}{1.000pt}}%
\put(440,649){\makebox(0,0)[r]{fit}}
\multiput(462,649)(20.756,0.000){4}{\usebox{\plotpoint}}
\put(528,649){\usebox{\plotpoint}}
\put(251.00,113.00){\usebox{\plotpoint}}
\put(255.44,133.25){\usebox{\plotpoint}}
\put(261.78,152.98){\usebox{\plotpoint}}
\put(269.03,172.42){\usebox{\plotpoint}}
\multiput(270,175)(8.176,19.077){0}{\usebox{\plotpoint}}
\put(277.33,191.43){\usebox{\plotpoint}}
\put(287.27,209.65){\usebox{\plotpoint}}
\multiput(288,211)(12.743,16.383){0}{\usebox{\plotpoint}}
\put(299.34,226.51){\usebox{\plotpoint}}
\multiput(301,229)(12.453,16.604){0}{\usebox{\plotpoint}}
\put(312.05,242.89){\usebox{\plotpoint}}
\multiput(313,244)(14.676,14.676){0}{\usebox{\plotpoint}}
\multiput(320,251)(13.508,15.759){0}{\usebox{\plotpoint}}
\put(326.11,258.13){\usebox{\plotpoint}}
\multiput(332,265)(14.676,14.676){0}{\usebox{\plotpoint}}
\put(340.10,273.45){\usebox{\plotpoint}}
\multiput(344,278)(15.759,13.508){0}{\usebox{\plotpoint}}
\put(354.92,287.92){\usebox{\plotpoint}}
\multiput(357,290)(15.945,13.287){0}{\usebox{\plotpoint}}
\multiput(363,295)(14.676,14.676){0}{\usebox{\plotpoint}}
\put(370.15,301.99){\usebox{\plotpoint}}
\multiput(376,307)(15.945,13.287){0}{\usebox{\plotpoint}}
\put(385.71,315.71){\usebox{\plotpoint}}
\multiput(388,318)(15.945,13.287){0}{\usebox{\plotpoint}}
\multiput(394,323)(15.945,13.287){0}{\usebox{\plotpoint}}
\put(401.44,329.23){\usebox{\plotpoint}}
\multiput(407,334)(15.945,13.287){0}{\usebox{\plotpoint}}
\put(417.32,342.60){\usebox{\plotpoint}}
\multiput(419,344)(15.945,13.287){0}{\usebox{\plotpoint}}
\multiput(425,349)(16.889,12.064){0}{\usebox{\plotpoint}}
\put(433.65,355.38){\usebox{\plotpoint}}
\multiput(438,359)(15.945,13.287){0}{\usebox{\plotpoint}}
\put(449.60,368.66){\usebox{\plotpoint}}
\multiput(450,369)(15.945,13.287){0}{\usebox{\plotpoint}}
\multiput(456,374)(16.889,12.064){0}{\usebox{\plotpoint}}
\put(465.93,381.44){\usebox{\plotpoint}}
\multiput(469,384)(15.945,13.287){0}{\usebox{\plotpoint}}
\multiput(475,389)(15.945,13.287){0}{\usebox{\plotpoint}}
\put(481.93,394.66){\usebox{\plotpoint}}
\multiput(488,399)(15.945,13.287){0}{\usebox{\plotpoint}}
\put(498.22,407.51){\usebox{\plotpoint}}
\multiput(500,409)(15.945,13.287){0}{\usebox{\plotpoint}}
\multiput(506,414)(17.270,11.513){0}{\usebox{\plotpoint}}
\put(514.78,419.98){\usebox{\plotpoint}}
\multiput(519,423)(15.945,13.287){0}{\usebox{\plotpoint}}
\put(530.96,432.96){\usebox{\plotpoint}}
\multiput(531,433)(15.945,13.287){0}{\usebox{\plotpoint}}
\multiput(537,438)(18.021,10.298){0}{\usebox{\plotpoint}}
\put(547.71,445.09){\usebox{\plotpoint}}
\multiput(550,447)(15.945,13.287){0}{\usebox{\plotpoint}}
\multiput(556,452)(15.945,13.287){0}{\usebox{\plotpoint}}
\put(563.79,458.19){\usebox{\plotpoint}}
\multiput(568,461)(16.889,12.064){0}{\usebox{\plotpoint}}
\put(580.45,470.54){\usebox{\plotpoint}}
\multiput(581,471)(15.945,13.287){0}{\usebox{\plotpoint}}
\multiput(587,476)(17.270,11.513){0}{\usebox{\plotpoint}}
\put(597.08,482.92){\usebox{\plotpoint}}
\multiput(600,485)(15.945,13.287){0}{\usebox{\plotpoint}}
\multiput(606,490)(17.270,11.513){0}{\usebox{\plotpoint}}
\put(613.65,495.38){\usebox{\plotpoint}}
\multiput(618,499)(15.945,13.287){0}{\usebox{\plotpoint}}
\put(629.93,508.23){\usebox{\plotpoint}}
\multiput(631,509)(17.270,11.513){0}{\usebox{\plotpoint}}
\multiput(637,513)(15.945,13.287){0}{\usebox{\plotpoint}}
\put(646.39,520.83){\usebox{\plotpoint}}
\multiput(649,523)(18.021,10.298){0}{\usebox{\plotpoint}}
\multiput(656,527)(15.945,13.287){0}{\usebox{\plotpoint}}
\put(663.14,532.95){\usebox{\plotpoint}}
\multiput(668,537)(17.270,11.513){0}{\usebox{\plotpoint}}
\put(679.55,545.62){\usebox{\plotpoint}}
\multiput(680,546)(16.889,12.064){0}{\usebox{\plotpoint}}
\multiput(687,551)(17.270,11.513){0}{\usebox{\plotpoint}}
\put(696.35,557.79){\usebox{\plotpoint}}
\multiput(699,560)(17.270,11.513){0}{\usebox{\plotpoint}}
\multiput(705,564)(16.889,12.064){0}{\usebox{\plotpoint}}
\put(713.14,569.95){\usebox{\plotpoint}}
\multiput(718,574)(17.270,11.513){0}{\usebox{\plotpoint}}
\put(729.55,582.62){\usebox{\plotpoint}}
\multiput(730,583)(15.945,13.287){0}{\usebox{\plotpoint}}
\multiput(736,588)(18.021,10.298){0}{\usebox{\plotpoint}}
\put(746.30,594.75){\usebox{\plotpoint}}
\multiput(749,597)(15.945,13.287){0}{\usebox{\plotpoint}}
\multiput(755,602)(17.270,11.513){0}{\usebox{\plotpoint}}
\put(762.80,607.29){\usebox{\plotpoint}}
\multiput(768,611)(17.270,11.513){0}{\usebox{\plotpoint}}
\put(779.50,619.58){\usebox{\plotpoint}}
\multiput(780,620)(15.945,13.287){0}{\usebox{\plotpoint}}
\multiput(786,625)(17.270,11.513){0}{\usebox{\plotpoint}}
\put(796.14,631.95){\usebox{\plotpoint}}
\multiput(799,634)(17.270,11.513){0}{\usebox{\plotpoint}}
\multiput(805,638)(15.945,13.287){0}{\usebox{\plotpoint}}
\put(812.70,644.42){\usebox{\plotpoint}}
\multiput(817,648)(18.021,10.298){0}{\usebox{\plotpoint}}
\put(829.45,656.54){\usebox{\plotpoint}}
\multiput(830,657)(15.945,13.287){0}{\usebox{\plotpoint}}
\put(836,662){\usebox{\plotpoint}}
\sbox{\plotpoint}{\rule[-0.200pt]{0.400pt}{0.400pt}}%
\put(440,604){\makebox(0,0)[r]{from $W_{ss}$}}
\put(484,604){\circle{24}}
\put(308,235){\circle{24}}
\put(344,282){\circle{24}}
\put(396,325){\circle{24}}
\put(417,344){\circle{24}}
\put(469,387){\circle{24}}
\put(484,400){\circle{24}}
\put(440,559){\makebox(0,0)[r]{from $W_{st}$}}
\put(484,559){\circle*{12}}
\put(308,238){\circle*{12}}
\put(344,282){\circle*{12}}
\put(372,310){\circle*{12}}
\put(396,324){\circle*{12}}
\put(417,344){\circle*{12}}
\put(469,387){\circle*{12}}
\put(484,401){\circle*{12}}
\put(484,397){\circle*{12}}
\put(572,464){\circle*{12}}
\put(660,529){\circle*{12}}
\put(748,566){\circle*{12}}
\put(462.0,559.0){\rule[-0.200pt]{15.899pt}{0.400pt}}
\put(462.0,549.0){\rule[-0.200pt]{0.400pt}{4.818pt}}
\put(528.0,549.0){\rule[-0.200pt]{0.400pt}{4.818pt}}
\put(308,238){\usebox{\plotpoint}}
\put(298.0,238.0){\rule[-0.200pt]{4.818pt}{0.400pt}}
\put(298.0,238.0){\rule[-0.200pt]{4.818pt}{0.400pt}}
\put(344,282){\usebox{\plotpoint}}
\put(334.0,282.0){\rule[-0.200pt]{4.818pt}{0.400pt}}
\put(334.0,282.0){\rule[-0.200pt]{4.818pt}{0.400pt}}
\put(372.0,310.0){\usebox{\plotpoint}}
\put(362.0,310.0){\rule[-0.200pt]{4.818pt}{0.400pt}}
\put(362.0,311.0){\rule[-0.200pt]{4.818pt}{0.400pt}}
\put(396.0,324.0){\usebox{\plotpoint}}
\put(386.0,324.0){\rule[-0.200pt]{4.818pt}{0.400pt}}
\put(386.0,325.0){\rule[-0.200pt]{4.818pt}{0.400pt}}
\put(417.0,344.0){\usebox{\plotpoint}}
\put(407.0,344.0){\rule[-0.200pt]{4.818pt}{0.400pt}}
\put(407.0,345.0){\rule[-0.200pt]{4.818pt}{0.400pt}}
\put(469.0,386.0){\rule[-0.200pt]{0.400pt}{0.482pt}}
\put(459.0,386.0){\rule[-0.200pt]{4.818pt}{0.400pt}}
\put(459.0,388.0){\rule[-0.200pt]{4.818pt}{0.400pt}}
\put(484.0,399.0){\rule[-0.200pt]{0.400pt}{0.723pt}}
\put(474.0,399.0){\rule[-0.200pt]{4.818pt}{0.400pt}}
\put(474.0,402.0){\rule[-0.200pt]{4.818pt}{0.400pt}}
\put(484.0,396.0){\usebox{\plotpoint}}
\put(474.0,396.0){\rule[-0.200pt]{4.818pt}{0.400pt}}
\put(474.0,397.0){\rule[-0.200pt]{4.818pt}{0.400pt}}
\put(572.0,462.0){\rule[-0.200pt]{0.400pt}{0.723pt}}
\put(562.0,462.0){\rule[-0.200pt]{4.818pt}{0.400pt}}
\put(562.0,465.0){\rule[-0.200pt]{4.818pt}{0.400pt}}
\put(660.0,523.0){\rule[-0.200pt]{0.400pt}{2.891pt}}
\put(650.0,523.0){\rule[-0.200pt]{4.818pt}{0.400pt}}
\put(650.0,535.0){\rule[-0.200pt]{4.818pt}{0.400pt}}
\put(748.0,538.0){\rule[-0.200pt]{0.400pt}{13.731pt}}
\put(738.0,538.0){\rule[-0.200pt]{4.818pt}{0.400pt}}
\put(738.0,595.0){\rule[-0.200pt]{4.818pt}{0.400pt}}
\end{picture}
\end{center}
\caption{Static potential computed on an anisotropic lattice with
$\eta\,a_t/a_s = 0.5$ and $\beta\!=\!1.8$. Potentials obtained from
spatial-temporal loops and from purely spatial loops are
shown. Results are preliminary.}
\label{fig:anisoV}
\end{figure}
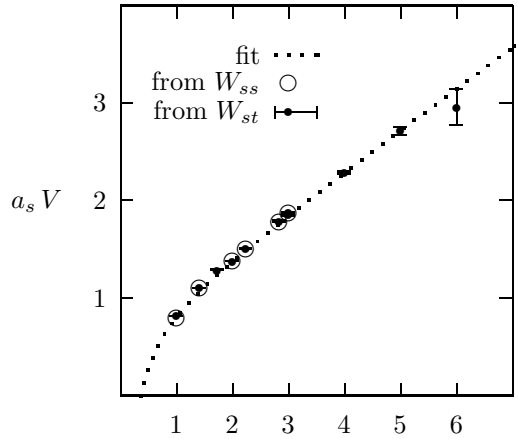

In Fig.~\ref{fig:anisoV} we show the static potential computed on an
anisotropic lattice from Wilson loops involving temporal and spatial
links, and from loops involving just spatial links. 
Shifted to account for renormalization, the two potentials agree well
for all~$r$, as they should.

Anisotropic lattices with $a_t\!\ll\!a_s$ are useful for any simulation
involving high-energy states, including glueballs, excited hadronic states, and
high-momentum final states in form factors. The dominant lattice errors in
relativistic heavy-quark physics are powers of $a_t M_Q$, where $M_Q$ is
the mass of the heavy quark; these are greatly reduced on highly anisotropic
lattices, which, therefore, provide an alternative to the NRQCD and
Fermilab approaches\,\cite{d234c}. 
A variety of quark and gluon actions designed
for anisotropic lattices are currently under development.

\subsection{Improved Gluons}
The plaquette-plus-rectangle gluon actions, Eqs.~(\ref{gluon-action})
and~(\ref{aniso-gluon-action}), have been used by several groups in
recent years for simulations of the static potential, QCD thermodynamics,
charmonium, the light-hadron spectrum, and so on. In every application area,
except perhaps the light-hadron spectrum, the improved action has led to
significantly improved results. 

One of the most striking recent applications
is in~\cite{morningstar-peardon}, a high-statistics
study of several different glueball masses at a wide range of lattice
spacings. What is striking is that this thorough study was carried out
entirely on desktop workstations, an achievement possible only through the
use of coarse, highly anisotropic (5:1) lattices. All the glueball masses
extrapolate to the values determined earlier with more traditional methods;
and all but one show relatively little dependence on the
lattice spacing, less than 10\%, out to lattice spacings of order 0.4\,fm.
Only the $0^{++}$~glueball shows strong $a$~dependence; its mass has a
20--25\% dip around 0.3\,fm. These are the largest measured finite-$a$ errors
in gluonic lattice QCD, and therefore the
$0^{++}$~glueball is of particular interest to designers of improved
gluon actions.  

A dip in the $0^{++}$~mass is also seen in SU(2)
QCD\,\cite{shakespeare-trottier}, although it is smaller, only about 14\%.
The authors of the SU(2) analysis noticed that the depth of the dip is reduced
almost by half when the Landau-gauge mean link~$u_0$ is used, in place of the
plaquette mean link~$u_p$, in the tadpole-improved tree-level gluon
action. This result is consistent with the observation that the
static-quark potential is more rotationally invariant with~$u_0$; both
argue that the correct order~$a^2$ correction in the action is stronger than
that obtained using~$u_p$. The SU(3) study used
$u_p$; it should be redone with~$u_0$ instead. Better still, we should
calculate the full $\order(a^2\alpha_V)$ corrections to the anisotropic gluon
action; work on this has begun. A small glueball radius would explain the
strong $a$~dependence of the $0^{++}$~mass; another explanation is that the
dip in the mass is caused by a phase transition nearby in the
fundamental-adjoint coupling constant plane of the gluon action.

\subsection{Improved Wilson Quarks}
Many of the talks at this meeting showed results for the SW quark action,
which is an order~$a$ improved version of the original Wilson lattice
action. Results are clearly superior when the SW action is
used instead of
the Wilson action, 
provided the clover coefficient is either tadpole improved or tuned
nonperturbatively; and generally SW and Wilson simulations extrapolate to
the same answer. Based on this work, the SW action should be accurate to
within a few percent for light-quark physics out to lattice spacings of
order~0.15--0.2\,fm. 

Larger lattice spacings require a more accurate action. A simple
construction that yields such an action begins with the continuum lagrangian:
\be
{\cal L} = \overline{\psi}_cM_c\psi_c, \quad\quad
M_c \equiv D\cdot\gamma + m_c.
\ee
Still working in the continuum, we introduce a field transformation,
\bearray
\psi_c &=& \Omega \psi \\
\Omega^2 &=& 1 - \half r a (D\cdot\gamma - m_c)
\eearray
to obtain a new lagrangian $\overline{\psi}M\psi$ where
\bearray
M &=& \Omega M_c \Omega \nl
  &=&  D\cdot\gamma  + m_c\left(1+\half r a
m_c\right) \\
  && -\half ra\left(D^2 + \half g\sigma\cdot F\right).
\eearray
The third term in $M$, introduced by the field transformation, breaks the
doubling symmetry when we discretize the action. Allowing errors only of
order~$a^4$ and higher, we obtain the D234c discretization
of~$M$\,\cite{d234c}:
\bearray
M_{\rm D234} &\equiv& \Delta\cdot\gamma - c_3 \,\frac{a^2}{6}
\Delta^{(3)}\cdot\gamma + m \\
&-& \frac{ra}{2}\left(\Delta^{(2)} + c_F \frac{g\sigma\cdot F}{2} 
-c_4 \frac{a^2}{12}\Delta^{(4)}\right) \nonumber
\eearray
where $m\equiv m_c (1+\half r a m_c)$, $\Delta^{(n)}_\mu$ is the lattice
discretization of the $n$'th gauge covariant derivative~$D^n_\mu$, and
$F_{\mu\nu}$ is an $a^2$-improved clover discretization of the gluon field
strength. All operators are tadpole improved; at tree level the couplings
$c_F=c_3=c_4=1$.

We have tested this action on lattices with $a=0.25$\,fm
and~0.4\,fm\,\cite{d234c}. We limited our study to quark masses near the
strange quark mass. Thus we avoid the inessential complications of chiral
extrapolation. Hadrons made of quarks with larger masses are generally
smaller and heavier; consequently discretization errors are probably larger
for the strange quark than for the other light quarks.

\begin{figure}
\begin{center}
\setlength{\unitlength}{0.240900pt}
\ifx\plotpoint\undefined\newsavebox{\plotpoint}\fi
\sbox{\plotpoint}{\rule[-0.200pt]{0.400pt}{0.400pt}}%
\begin{picture}(900,624)(0,125)
\font\gnuplot=cmr10 at 10pt
\gnuplot
\sbox{\plotpoint}{\rule[-0.200pt]{0.400pt}{0.400pt}}%
\put(220.0,113.0){\rule[-0.200pt]{148.394pt}{0.400pt}}
\put(220.0,113.0){\rule[-0.200pt]{0.400pt}{147.672pt}}
\put(220.0,207.0){\rule[-0.200pt]{4.818pt}{0.400pt}}
\put(198,207){\makebox(0,0)[r]{0.2}}
\put(816.0,207.0){\rule[-0.200pt]{4.818pt}{0.400pt}}
\put(220.0,302.0){\rule[-0.200pt]{4.818pt}{0.400pt}}
\put(198,302){\makebox(0,0)[r]{0.4}}
\put(816.0,302.0){\rule[-0.200pt]{4.818pt}{0.400pt}}
\put(220.0,396.0){\rule[-0.200pt]{4.818pt}{0.400pt}}
\put(198,396){\makebox(0,0)[r]{0.6}}
\put(816.0,396.0){\rule[-0.200pt]{4.818pt}{0.400pt}}
\put(220.0,490.0){\rule[-0.200pt]{4.818pt}{0.400pt}}
\put(198,490){\makebox(0,0)[r]{0.8}}
\put(816.0,490.0){\rule[-0.200pt]{4.818pt}{0.400pt}}
\put(220.0,585.0){\rule[-0.200pt]{4.818pt}{0.400pt}}
\put(198,585){\makebox(0,0)[r]{1.0}}
\put(816.0,585.0){\rule[-0.200pt]{4.818pt}{0.400pt}}
\put(220.0,679.0){\rule[-0.200pt]{4.818pt}{0.400pt}}
\put(198,679){\makebox(0,0)[r]{1.2}}
\put(816.0,679.0){\rule[-0.200pt]{4.818pt}{0.400pt}}
\put(360.0,113.0){\rule[-0.200pt]{0.400pt}{4.818pt}}
\put(360,68){\makebox(0,0){0.5}}
\put(360.0,706.0){\rule[-0.200pt]{0.400pt}{4.818pt}}
\put(500.0,113.0){\rule[-0.200pt]{0.400pt}{4.818pt}}
\put(500,68){\makebox(0,0){1.0}}
\put(500.0,706.0){\rule[-0.200pt]{0.400pt}{4.818pt}}
\put(640.0,113.0){\rule[-0.200pt]{0.400pt}{4.818pt}}
\put(640,68){\makebox(0,0){1.5}}
\put(640.0,706.0){\rule[-0.200pt]{0.400pt}{4.818pt}}
\put(780.0,113.0){\rule[-0.200pt]{0.400pt}{4.818pt}}
\put(780,68){\makebox(0,0){2.0}}
\put(780.0,706.0){\rule[-0.200pt]{0.400pt}{4.818pt}}
\put(220.0,113.0){\rule[-0.200pt]{148.394pt}{0.400pt}}
\put(836.0,113.0){\rule[-0.200pt]{0.400pt}{147.672pt}}
\put(220.0,726.0){\rule[-0.200pt]{148.394pt}{0.400pt}}
\put(45,419){\makebox(0,0){$c^2(p)$}}
\put(550,23){\makebox(0,0){$p a$ }}
\put(724,585){\makebox(0,0)[l]{$\pi_s$}}
\put(724,639){\makebox(0,0)[l]{$\phi$}}
\put(724,420){\makebox(0,0)[l]{$\pi_s$}}
\put(724,207){\makebox(0,0)[l]{$\phi$}}
\put(248,632){\makebox(0,0)[l]{D234c}}
\put(248,443){\makebox(0,0)[l]{SW}}
\put(220.0,113.0){\rule[-0.200pt]{0.400pt}{147.672pt}}
\put(440,566){\circle*{18}}
\put(572,552){\circle*{18}}
\put(635,552){\circle*{18}}
\put(660,537){\circle*{18}}
\put(718,537){\circle*{18}}
\put(764,509){\circle*{18}}
\put(440.0,556.0){\rule[-0.200pt]{0.400pt}{4.577pt}}
\put(430.0,556.0){\rule[-0.200pt]{4.818pt}{0.400pt}}
\put(430.0,575.0){\rule[-0.200pt]{4.818pt}{0.400pt}}
\put(572.0,537.0){\rule[-0.200pt]{0.400pt}{6.986pt}}
\put(562.0,537.0){\rule[-0.200pt]{4.818pt}{0.400pt}}
\put(562.0,566.0){\rule[-0.200pt]{4.818pt}{0.400pt}}
\put(635.0,537.0){\rule[-0.200pt]{0.400pt}{6.986pt}}
\put(625.0,537.0){\rule[-0.200pt]{4.818pt}{0.400pt}}
\put(625.0,566.0){\rule[-0.200pt]{4.818pt}{0.400pt}}
\put(660.0,528.0){\rule[-0.200pt]{0.400pt}{4.577pt}}
\put(650.0,528.0){\rule[-0.200pt]{4.818pt}{0.400pt}}
\put(650.0,547.0){\rule[-0.200pt]{4.818pt}{0.400pt}}
\put(718.0,523.0){\rule[-0.200pt]{0.400pt}{6.986pt}}
\put(708.0,523.0){\rule[-0.200pt]{4.818pt}{0.400pt}}
\put(708.0,552.0){\rule[-0.200pt]{4.818pt}{0.400pt}}
\put(764.0,486.0){\rule[-0.200pt]{0.400pt}{11.322pt}}
\put(754.0,486.0){\rule[-0.200pt]{4.818pt}{0.400pt}}
\put(754.0,533.0){\rule[-0.200pt]{4.818pt}{0.400pt}}
\put(440,608){\circle{18}}
\put(572,622){\circle{18}}
\put(635,665){\circle{18}}
\put(718,717){\circle{18}}
\put(440.0,585.0){\rule[-0.200pt]{0.400pt}{11.322pt}}
\put(430.0,585.0){\rule[-0.200pt]{4.818pt}{0.400pt}}
\put(430.0,632.0){\rule[-0.200pt]{4.818pt}{0.400pt}}
\put(572.0,608.0){\rule[-0.200pt]{0.400pt}{6.745pt}}
\put(562.0,608.0){\rule[-0.200pt]{4.818pt}{0.400pt}}
\put(562.0,636.0){\rule[-0.200pt]{4.818pt}{0.400pt}}
\put(635.0,646.0){\rule[-0.200pt]{0.400pt}{9.154pt}}
\put(625.0,646.0){\rule[-0.200pt]{4.818pt}{0.400pt}}
\put(625.0,684.0){\rule[-0.200pt]{4.818pt}{0.400pt}}
\put(718.0,674.0){\rule[-0.200pt]{0.400pt}{12.527pt}}
\put(708.0,674.0){\rule[-0.200pt]{4.818pt}{0.400pt}}
\put(708.0,726.0){\rule[-0.200pt]{4.818pt}{0.400pt}}
\put(440,411){\circle*{18}}
\put(572,373){\circle*{18}}
\put(635,377){\circle*{18}}
\put(660,353){\circle*{18}}
\put(718,356){\circle*{18}}
\put(764,351){\circle*{18}}
\put(440.0,400.0){\rule[-0.200pt]{0.400pt}{5.300pt}}
\put(430.0,400.0){\rule[-0.200pt]{4.818pt}{0.400pt}}
\put(430.0,422.0){\rule[-0.200pt]{4.818pt}{0.400pt}}
\put(572.0,367.0){\rule[-0.200pt]{0.400pt}{2.891pt}}
\put(562.0,367.0){\rule[-0.200pt]{4.818pt}{0.400pt}}
\put(562.0,379.0){\rule[-0.200pt]{4.818pt}{0.400pt}}
\put(635.0,370.0){\rule[-0.200pt]{0.400pt}{3.373pt}}
\put(625.0,370.0){\rule[-0.200pt]{4.818pt}{0.400pt}}
\put(625.0,384.0){\rule[-0.200pt]{4.818pt}{0.400pt}}
\put(660.0,346.0){\rule[-0.200pt]{0.400pt}{3.613pt}}
\put(650.0,346.0){\rule[-0.200pt]{4.818pt}{0.400pt}}
\put(650.0,361.0){\rule[-0.200pt]{4.818pt}{0.400pt}}
\put(718.0,347.0){\rule[-0.200pt]{0.400pt}{4.336pt}}
\put(708.0,347.0){\rule[-0.200pt]{4.818pt}{0.400pt}}
\put(708.0,365.0){\rule[-0.200pt]{4.818pt}{0.400pt}}
\put(764.0,325.0){\rule[-0.200pt]{0.400pt}{12.527pt}}
\put(754.0,325.0){\rule[-0.200pt]{4.818pt}{0.400pt}}
\put(754.0,377.0){\rule[-0.200pt]{4.818pt}{0.400pt}}
\put(440,312){\circle{18}}
\put(572,274){\circle{18}}
\put(635,280){\circle{18}}
\put(718,276){\circle{18}}
\put(440.0,304.0){\rule[-0.200pt]{0.400pt}{3.613pt}}
\put(430.0,304.0){\rule[-0.200pt]{4.818pt}{0.400pt}}
\put(430.0,319.0){\rule[-0.200pt]{4.818pt}{0.400pt}}
\put(572.0,268.0){\rule[-0.200pt]{0.400pt}{2.891pt}}
\put(562.0,268.0){\rule[-0.200pt]{4.818pt}{0.400pt}}
\put(562.0,280.0){\rule[-0.200pt]{4.818pt}{0.400pt}}
\put(635.0,274.0){\rule[-0.200pt]{0.400pt}{3.132pt}}
\put(625.0,274.0){\rule[-0.200pt]{4.818pt}{0.400pt}}
\put(625.0,287.0){\rule[-0.200pt]{4.818pt}{0.400pt}}
\put(718.0,270.0){\rule[-0.200pt]{0.400pt}{2.891pt}}
\put(708.0,270.0){\rule[-0.200pt]{4.818pt}{0.400pt}}
\put(708.0,282.0){\rule[-0.200pt]{4.818pt}{0.400pt}}
\sbox{\plotpoint}{\rule[-0.500pt]{1.000pt}{1.000pt}}%
\put(220,565){\usebox{\plotpoint}}
\put(220.00,565.00){\usebox{\plotpoint}}
\multiput(226,565)(20.756,0.000){0}{\usebox{\plotpoint}}
\multiput(232,565)(20.756,0.000){0}{\usebox{\plotpoint}}
\put(240.76,565.00){\usebox{\plotpoint}}
\multiput(245,565)(20.756,0.000){0}{\usebox{\plotpoint}}
\multiput(251,565)(20.756,0.000){0}{\usebox{\plotpoint}}
\put(261.51,565.00){\usebox{\plotpoint}}
\multiput(264,565)(20.756,0.000){0}{\usebox{\plotpoint}}
\multiput(270,565)(20.756,0.000){0}{\usebox{\plotpoint}}
\multiput(276,565)(20.756,0.000){0}{\usebox{\plotpoint}}
\put(282.27,565.00){\usebox{\plotpoint}}
\multiput(288,565)(20.756,0.000){0}{\usebox{\plotpoint}}
\multiput(295,565)(20.756,0.000){0}{\usebox{\plotpoint}}
\put(303.02,565.00){\usebox{\plotpoint}}
\multiput(307,565)(20.756,0.000){0}{\usebox{\plotpoint}}
\multiput(313,565)(20.756,0.000){0}{\usebox{\plotpoint}}
\put(323.78,565.00){\usebox{\plotpoint}}
\multiput(326,565)(20.756,0.000){0}{\usebox{\plotpoint}}
\multiput(332,565)(20.756,0.000){0}{\usebox{\plotpoint}}
\multiput(338,565)(20.756,0.000){0}{\usebox{\plotpoint}}
\put(344.53,565.00){\usebox{\plotpoint}}
\multiput(351,565)(20.756,0.000){0}{\usebox{\plotpoint}}
\multiput(357,565)(20.473,-3.412){0}{\usebox{\plotpoint}}
\put(365.21,564.00){\usebox{\plotpoint}}
\multiput(369,564)(20.756,0.000){0}{\usebox{\plotpoint}}
\multiput(376,564)(20.756,0.000){0}{\usebox{\plotpoint}}
\put(385.96,564.00){\usebox{\plotpoint}}
\multiput(388,564)(20.756,0.000){0}{\usebox{\plotpoint}}
\multiput(394,564)(20.756,0.000){0}{\usebox{\plotpoint}}
\put(406.72,564.00){\usebox{\plotpoint}}
\multiput(407,564)(20.756,0.000){0}{\usebox{\plotpoint}}
\multiput(413,564)(20.756,0.000){0}{\usebox{\plotpoint}}
\multiput(419,564)(20.756,0.000){0}{\usebox{\plotpoint}}
\put(427.47,564.00){\usebox{\plotpoint}}
\multiput(432,564)(20.473,-3.412){0}{\usebox{\plotpoint}}
\multiput(438,563)(20.756,0.000){0}{\usebox{\plotpoint}}
\put(448.14,563.00){\usebox{\plotpoint}}
\multiput(450,563)(20.756,0.000){0}{\usebox{\plotpoint}}
\multiput(456,563)(20.756,0.000){0}{\usebox{\plotpoint}}
\put(468.82,562.03){\usebox{\plotpoint}}
\multiput(469,562)(20.756,0.000){0}{\usebox{\plotpoint}}
\multiput(475,562)(20.756,0.000){0}{\usebox{\plotpoint}}
\multiput(481,562)(20.756,0.000){0}{\usebox{\plotpoint}}
\put(489.55,561.74){\usebox{\plotpoint}}
\multiput(494,561)(20.756,0.000){0}{\usebox{\plotpoint}}
\multiput(500,561)(20.756,0.000){0}{\usebox{\plotpoint}}
\put(510.19,560.30){\usebox{\plotpoint}}
\multiput(512,560)(20.756,0.000){0}{\usebox{\plotpoint}}
\multiput(519,560)(20.473,-3.412){0}{\usebox{\plotpoint}}
\put(530.84,559.00){\usebox{\plotpoint}}
\multiput(531,559)(20.756,0.000){0}{\usebox{\plotpoint}}
\multiput(537,559)(20.547,-2.935){0}{\usebox{\plotpoint}}
\multiput(544,558)(20.473,-3.412){0}{\usebox{\plotpoint}}
\put(551.44,557.00){\usebox{\plotpoint}}
\multiput(556,557)(20.473,-3.412){0}{\usebox{\plotpoint}}
\multiput(562,556)(20.756,0.000){0}{\usebox{\plotpoint}}
\put(572.07,555.42){\usebox{\plotpoint}}
\multiput(575,555)(20.473,-3.412){0}{\usebox{\plotpoint}}
\multiput(581,554)(20.756,0.000){0}{\usebox{\plotpoint}}
\put(592.63,553.06){\usebox{\plotpoint}}
\multiput(593,553)(20.547,-2.935){0}{\usebox{\plotpoint}}
\multiput(600,552)(20.473,-3.412){0}{\usebox{\plotpoint}}
\multiput(606,551)(20.473,-3.412){0}{\usebox{\plotpoint}}
\put(613.13,549.81){\usebox{\plotpoint}}
\multiput(618,549)(20.473,-3.412){0}{\usebox{\plotpoint}}
\multiput(624,548)(20.547,-2.935){0}{\usebox{\plotpoint}}
\put(633.63,546.56){\usebox{\plotpoint}}
\multiput(637,546)(20.473,-3.412){0}{\usebox{\plotpoint}}
\multiput(643,545)(20.473,-3.412){0}{\usebox{\plotpoint}}
\put(654.12,543.27){\usebox{\plotpoint}}
\multiput(656,543)(19.690,-6.563){0}{\usebox{\plotpoint}}
\multiput(662,541)(20.473,-3.412){0}{\usebox{\plotpoint}}
\multiput(668,540)(20.473,-3.412){0}{\usebox{\plotpoint}}
\put(674.35,538.88){\usebox{\plotpoint}}
\multiput(680,537)(20.547,-2.935){0}{\usebox{\plotpoint}}
\multiput(687,536)(19.690,-6.563){0}{\usebox{\plotpoint}}
\put(694.33,533.56){\usebox{\plotpoint}}
\multiput(699,532)(20.473,-3.412){0}{\usebox{\plotpoint}}
\multiput(705,531)(19.957,-5.702){0}{\usebox{\plotpoint}}
\put(714.35,528.22){\usebox{\plotpoint}}
\multiput(718,527)(19.690,-6.563){0}{\usebox{\plotpoint}}
\multiput(724,525)(19.690,-6.563){0}{\usebox{\plotpoint}}
\put(734.04,521.65){\usebox{\plotpoint}}
\multiput(736,521)(19.957,-5.702){0}{\usebox{\plotpoint}}
\multiput(743,519)(19.690,-6.563){0}{\usebox{\plotpoint}}
\put(753.54,514.73){\usebox{\plotpoint}}
\multiput(755,514)(19.690,-6.563){0}{\usebox{\plotpoint}}
\multiput(761,512)(19.957,-5.702){0}{\usebox{\plotpoint}}
\put(772.94,507.53){\usebox{\plotpoint}}
\multiput(774,507)(18.564,-9.282){0}{\usebox{\plotpoint}}
\multiput(780,504)(19.690,-6.563){0}{\usebox{\plotpoint}}
\put(791.85,499.08){\usebox{\plotpoint}}
\multiput(792,499)(19.077,-8.176){0}{\usebox{\plotpoint}}
\multiput(799,496)(18.564,-9.282){0}{\usebox{\plotpoint}}
\put(810.60,490.20){\usebox{\plotpoint}}
\multiput(811,490)(18.564,-9.282){0}{\usebox{\plotpoint}}
\multiput(817,487)(18.021,-10.298){0}{\usebox{\plotpoint}}
\put(828.95,480.52){\usebox{\plotpoint}}
\multiput(830,480)(17.270,-11.513){0}{\usebox{\plotpoint}}
\put(836,476){\usebox{\plotpoint}}
\put(220,597){\usebox{\plotpoint}}
\put(220.00,597.00){\usebox{\plotpoint}}
\multiput(226,597)(20.756,0.000){0}{\usebox{\plotpoint}}
\multiput(232,597)(20.756,0.000){0}{\usebox{\plotpoint}}
\put(240.76,597.00){\usebox{\plotpoint}}
\multiput(245,597)(20.756,0.000){0}{\usebox{\plotpoint}}
\multiput(251,597)(20.756,0.000){0}{\usebox{\plotpoint}}
\put(261.51,597.00){\usebox{\plotpoint}}
\multiput(264,597)(20.756,0.000){0}{\usebox{\plotpoint}}
\multiput(270,597)(20.756,0.000){0}{\usebox{\plotpoint}}
\multiput(276,597)(20.756,0.000){0}{\usebox{\plotpoint}}
\put(282.27,597.00){\usebox{\plotpoint}}
\multiput(288,597)(20.756,0.000){0}{\usebox{\plotpoint}}
\multiput(295,597)(20.756,0.000){0}{\usebox{\plotpoint}}
\put(303.02,597.00){\usebox{\plotpoint}}
\multiput(307,597)(20.756,0.000){0}{\usebox{\plotpoint}}
\multiput(313,597)(20.756,0.000){0}{\usebox{\plotpoint}}
\put(323.78,597.00){\usebox{\plotpoint}}
\multiput(326,597)(20.756,0.000){0}{\usebox{\plotpoint}}
\multiput(332,597)(20.756,0.000){0}{\usebox{\plotpoint}}
\multiput(338,597)(20.756,0.000){0}{\usebox{\plotpoint}}
\put(344.53,597.00){\usebox{\plotpoint}}
\multiput(351,597)(20.473,3.412){0}{\usebox{\plotpoint}}
\multiput(357,598)(20.756,0.000){0}{\usebox{\plotpoint}}
\put(365.21,598.00){\usebox{\plotpoint}}
\multiput(369,598)(20.756,0.000){0}{\usebox{\plotpoint}}
\multiput(376,598)(20.756,0.000){0}{\usebox{\plotpoint}}
\put(385.96,598.00){\usebox{\plotpoint}}
\multiput(388,598)(20.473,3.412){0}{\usebox{\plotpoint}}
\multiput(394,599)(20.756,0.000){0}{\usebox{\plotpoint}}
\put(406.63,599.00){\usebox{\plotpoint}}
\multiput(407,599)(20.473,3.412){0}{\usebox{\plotpoint}}
\multiput(413,600)(20.756,0.000){0}{\usebox{\plotpoint}}
\multiput(419,600)(20.756,0.000){0}{\usebox{\plotpoint}}
\put(427.28,600.33){\usebox{\plotpoint}}
\multiput(432,601)(20.756,0.000){0}{\usebox{\plotpoint}}
\multiput(438,601)(20.473,3.412){0}{\usebox{\plotpoint}}
\put(447.86,602.64){\usebox{\plotpoint}}
\multiput(450,603)(20.756,0.000){0}{\usebox{\plotpoint}}
\multiput(456,603)(20.547,2.935){0}{\usebox{\plotpoint}}
\put(468.44,604.91){\usebox{\plotpoint}}
\multiput(469,605)(20.473,3.412){0}{\usebox{\plotpoint}}
\multiput(475,606)(20.756,0.000){0}{\usebox{\plotpoint}}
\multiput(481,606)(20.547,2.935){0}{\usebox{\plotpoint}}
\put(489.01,607.17){\usebox{\plotpoint}}
\multiput(494,608)(19.690,6.563){0}{\usebox{\plotpoint}}
\multiput(500,610)(20.473,3.412){0}{\usebox{\plotpoint}}
\put(509.25,611.54){\usebox{\plotpoint}}
\multiput(512,612)(20.547,2.935){0}{\usebox{\plotpoint}}
\multiput(519,613)(19.690,6.563){0}{\usebox{\plotpoint}}
\put(529.51,615.75){\usebox{\plotpoint}}
\multiput(531,616)(19.690,6.563){0}{\usebox{\plotpoint}}
\multiput(537,618)(20.547,2.935){0}{\usebox{\plotpoint}}
\put(549.55,620.85){\usebox{\plotpoint}}
\multiput(550,621)(19.690,6.563){0}{\usebox{\plotpoint}}
\multiput(556,623)(19.690,6.563){0}{\usebox{\plotpoint}}
\multiput(562,625)(19.690,6.563){0}{\usebox{\plotpoint}}
\put(569.20,627.51){\usebox{\plotpoint}}
\multiput(575,630)(19.690,6.563){0}{\usebox{\plotpoint}}
\multiput(581,632)(19.690,6.563){0}{\usebox{\plotpoint}}
\put(588.61,634.80){\usebox{\plotpoint}}
\multiput(593,637)(19.077,8.176){0}{\usebox{\plotpoint}}
\multiput(600,640)(18.564,9.282){0}{\usebox{\plotpoint}}
\put(607.36,643.68){\usebox{\plotpoint}}
\multiput(612,646)(18.564,9.282){0}{\usebox{\plotpoint}}
\multiput(618,649)(18.564,9.282){0}{\usebox{\plotpoint}}
\put(625.87,653.07){\usebox{\plotpoint}}
\multiput(631,656)(18.564,9.282){0}{\usebox{\plotpoint}}
\multiput(637,659)(17.270,11.513){0}{\usebox{\plotpoint}}
\put(643.77,663.51){\usebox{\plotpoint}}
\multiput(649,667)(18.021,10.298){0}{\usebox{\plotpoint}}
\put(660.92,675.10){\usebox{\plotpoint}}
\multiput(662,676)(17.270,11.513){0}{\usebox{\plotpoint}}
\multiput(668,680)(15.945,13.287){0}{\usebox{\plotpoint}}
\put(677.33,687.77){\usebox{\plotpoint}}
\multiput(680,690)(16.889,12.064){0}{\usebox{\plotpoint}}
\multiput(687,695)(15.945,13.287){0}{\usebox{\plotpoint}}
\put(693.61,700.61){\usebox{\plotpoint}}
\multiput(699,706)(14.676,14.676){0}{\usebox{\plotpoint}}
\put(708.53,715.03){\usebox{\plotpoint}}
\multiput(712,718)(14.676,14.676){0}{\usebox{\plotpoint}}
\multiput(718,724)(14.676,14.676){0}{\usebox{\plotpoint}}
\put(720,726){\usebox{\plotpoint}}
\put(220,410){\usebox{\plotpoint}}
\put(220.00,410.00){\usebox{\plotpoint}}
\multiput(226,410)(20.756,0.000){0}{\usebox{\plotpoint}}
\multiput(232,410)(20.756,0.000){0}{\usebox{\plotpoint}}
\put(240.76,410.00){\usebox{\plotpoint}}
\multiput(245,410)(20.756,0.000){0}{\usebox{\plotpoint}}
\multiput(251,410)(20.756,0.000){0}{\usebox{\plotpoint}}
\put(261.51,410.00){\usebox{\plotpoint}}
\multiput(264,410)(20.473,-3.412){0}{\usebox{\plotpoint}}
\multiput(270,409)(20.756,0.000){0}{\usebox{\plotpoint}}
\multiput(276,409)(20.756,0.000){0}{\usebox{\plotpoint}}
\put(282.18,409.00){\usebox{\plotpoint}}
\multiput(288,409)(20.756,0.000){0}{\usebox{\plotpoint}}
\multiput(295,409)(20.473,-3.412){0}{\usebox{\plotpoint}}
\put(302.86,408.00){\usebox{\plotpoint}}
\multiput(307,408)(20.756,0.000){0}{\usebox{\plotpoint}}
\multiput(313,408)(20.756,0.000){0}{\usebox{\plotpoint}}
\put(323.56,407.41){\usebox{\plotpoint}}
\multiput(326,407)(20.756,0.000){0}{\usebox{\plotpoint}}
\multiput(332,407)(20.756,0.000){0}{\usebox{\plotpoint}}
\multiput(338,407)(20.473,-3.412){0}{\usebox{\plotpoint}}
\put(344.20,406.00){\usebox{\plotpoint}}
\multiput(351,406)(20.756,0.000){0}{\usebox{\plotpoint}}
\multiput(357,406)(20.473,-3.412){0}{\usebox{\plotpoint}}
\put(364.87,405.00){\usebox{\plotpoint}}
\multiput(369,405)(20.547,-2.935){0}{\usebox{\plotpoint}}
\multiput(376,404)(20.756,0.000){0}{\usebox{\plotpoint}}
\put(385.51,403.41){\usebox{\plotpoint}}
\multiput(388,403)(20.756,0.000){0}{\usebox{\plotpoint}}
\multiput(394,403)(20.473,-3.412){0}{\usebox{\plotpoint}}
\put(406.15,402.00){\usebox{\plotpoint}}
\multiput(407,402)(20.473,-3.412){0}{\usebox{\plotpoint}}
\multiput(413,401)(20.756,0.000){0}{\usebox{\plotpoint}}
\multiput(419,401)(20.473,-3.412){0}{\usebox{\plotpoint}}
\put(426.72,399.75){\usebox{\plotpoint}}
\multiput(432,399)(20.756,0.000){0}{\usebox{\plotpoint}}
\multiput(438,399)(20.473,-3.412){0}{\usebox{\plotpoint}}
\put(447.30,397.45){\usebox{\plotpoint}}
\multiput(450,397)(20.756,0.000){0}{\usebox{\plotpoint}}
\multiput(456,397)(20.547,-2.935){0}{\usebox{\plotpoint}}
\put(467.88,395.19){\usebox{\plotpoint}}
\multiput(469,395)(20.473,-3.412){0}{\usebox{\plotpoint}}
\multiput(475,394)(20.756,0.000){0}{\usebox{\plotpoint}}
\multiput(481,394)(20.547,-2.935){0}{\usebox{\plotpoint}}
\put(488.46,392.92){\usebox{\plotpoint}}
\multiput(494,392)(20.473,-3.412){0}{\usebox{\plotpoint}}
\multiput(500,391)(20.473,-3.412){0}{\usebox{\plotpoint}}
\put(508.93,389.51){\usebox{\plotpoint}}
\multiput(512,389)(20.756,0.000){0}{\usebox{\plotpoint}}
\multiput(519,389)(20.473,-3.412){0}{\usebox{\plotpoint}}
\put(529.50,387.25){\usebox{\plotpoint}}
\multiput(531,387)(20.473,-3.412){0}{\usebox{\plotpoint}}
\multiput(537,386)(20.547,-2.935){0}{\usebox{\plotpoint}}
\put(549.99,384.00){\usebox{\plotpoint}}
\multiput(550,384)(20.473,-3.412){0}{\usebox{\plotpoint}}
\multiput(556,383)(20.473,-3.412){0}{\usebox{\plotpoint}}
\multiput(562,382)(20.473,-3.412){0}{\usebox{\plotpoint}}
\put(570.48,380.65){\usebox{\plotpoint}}
\multiput(575,380)(20.473,-3.412){0}{\usebox{\plotpoint}}
\multiput(581,379)(20.473,-3.412){0}{\usebox{\plotpoint}}
\put(590.97,377.34){\usebox{\plotpoint}}
\multiput(593,377)(19.957,-5.702){0}{\usebox{\plotpoint}}
\multiput(600,375)(20.473,-3.412){0}{\usebox{\plotpoint}}
\put(611.26,373.12){\usebox{\plotpoint}}
\multiput(612,373)(20.473,-3.412){0}{\usebox{\plotpoint}}
\multiput(618,372)(20.473,-3.412){0}{\usebox{\plotpoint}}
\multiput(624,371)(19.957,-5.702){0}{\usebox{\plotpoint}}
\put(631.55,368.91){\usebox{\plotpoint}}
\multiput(637,368)(20.473,-3.412){0}{\usebox{\plotpoint}}
\multiput(643,367)(20.473,-3.412){0}{\usebox{\plotpoint}}
\put(651.95,365.16){\usebox{\plotpoint}}
\multiput(656,364)(20.473,-3.412){0}{\usebox{\plotpoint}}
\multiput(662,363)(20.473,-3.412){0}{\usebox{\plotpoint}}
\put(672.15,360.62){\usebox{\plotpoint}}
\multiput(674,360)(20.473,-3.412){0}{\usebox{\plotpoint}}
\multiput(680,359)(20.547,-2.935){0}{\usebox{\plotpoint}}
\put(692.36,356.21){\usebox{\plotpoint}}
\multiput(693,356)(20.473,-3.412){0}{\usebox{\plotpoint}}
\multiput(699,355)(19.690,-6.563){0}{\usebox{\plotpoint}}
\multiput(705,353)(20.547,-2.935){0}{\usebox{\plotpoint}}
\put(712.57,351.81){\usebox{\plotpoint}}
\multiput(718,350)(20.473,-3.412){0}{\usebox{\plotpoint}}
\multiput(724,349)(19.690,-6.563){0}{\usebox{\plotpoint}}
\put(732.59,346.57){\usebox{\plotpoint}}
\multiput(736,346)(19.957,-5.702){0}{\usebox{\plotpoint}}
\multiput(743,344)(20.473,-3.412){0}{\usebox{\plotpoint}}
\put(752.74,341.75){\usebox{\plotpoint}}
\multiput(755,341)(20.473,-3.412){0}{\usebox{\plotpoint}}
\multiput(761,340)(19.957,-5.702){0}{\usebox{\plotpoint}}
\put(772.75,336.42){\usebox{\plotpoint}}
\multiput(774,336)(20.473,-3.412){0}{\usebox{\plotpoint}}
\multiput(780,335)(19.690,-6.563){0}{\usebox{\plotpoint}}
\multiput(786,333)(19.690,-6.563){0}{\usebox{\plotpoint}}
\put(792.70,330.90){\usebox{\plotpoint}}
\multiput(799,330)(19.690,-6.563){0}{\usebox{\plotpoint}}
\multiput(805,328)(19.690,-6.563){0}{\usebox{\plotpoint}}
\put(812.65,325.45){\usebox{\plotpoint}}
\multiput(817,324)(19.957,-5.702){0}{\usebox{\plotpoint}}
\multiput(824,322)(20.473,-3.412){0}{\usebox{\plotpoint}}
\put(832.66,320.11){\usebox{\plotpoint}}
\put(836,319){\usebox{\plotpoint}}
\put(220,306){\usebox{\plotpoint}}
\put(220.00,306.00){\usebox{\plotpoint}}
\multiput(226,306)(20.756,0.000){0}{\usebox{\plotpoint}}
\multiput(232,306)(20.756,0.000){0}{\usebox{\plotpoint}}
\put(240.76,306.00){\usebox{\plotpoint}}
\multiput(245,306)(20.756,0.000){0}{\usebox{\plotpoint}}
\multiput(251,306)(20.756,0.000){0}{\usebox{\plotpoint}}
\put(261.51,306.00){\usebox{\plotpoint}}
\multiput(264,306)(20.756,0.000){0}{\usebox{\plotpoint}}
\multiput(270,306)(20.756,0.000){0}{\usebox{\plotpoint}}
\multiput(276,306)(20.756,0.000){0}{\usebox{\plotpoint}}
\put(282.27,306.00){\usebox{\plotpoint}}
\multiput(288,306)(20.756,0.000){0}{\usebox{\plotpoint}}
\multiput(295,306)(20.756,0.000){0}{\usebox{\plotpoint}}
\put(302.99,305.67){\usebox{\plotpoint}}
\multiput(307,305)(20.756,0.000){0}{\usebox{\plotpoint}}
\multiput(313,305)(20.756,0.000){0}{\usebox{\plotpoint}}
\put(323.69,305.00){\usebox{\plotpoint}}
\multiput(326,305)(20.756,0.000){0}{\usebox{\plotpoint}}
\multiput(332,305)(20.756,0.000){0}{\usebox{\plotpoint}}
\multiput(338,305)(20.473,-3.412){0}{\usebox{\plotpoint}}
\put(344.37,304.00){\usebox{\plotpoint}}
\multiput(351,304)(20.756,0.000){0}{\usebox{\plotpoint}}
\multiput(357,304)(20.756,0.000){0}{\usebox{\plotpoint}}
\put(365.12,304.00){\usebox{\plotpoint}}
\multiput(369,304)(20.547,-2.935){0}{\usebox{\plotpoint}}
\multiput(376,303)(20.756,0.000){0}{\usebox{\plotpoint}}
\put(385.81,303.00){\usebox{\plotpoint}}
\multiput(388,303)(20.756,0.000){0}{\usebox{\plotpoint}}
\multiput(394,303)(20.473,-3.412){0}{\usebox{\plotpoint}}
\put(406.48,302.00){\usebox{\plotpoint}}
\multiput(407,302)(20.756,0.000){0}{\usebox{\plotpoint}}
\multiput(413,302)(20.756,0.000){0}{\usebox{\plotpoint}}
\multiput(419,302)(20.473,-3.412){0}{\usebox{\plotpoint}}
\put(427.15,301.00){\usebox{\plotpoint}}
\multiput(432,301)(20.756,0.000){0}{\usebox{\plotpoint}}
\multiput(438,301)(20.473,-3.412){0}{\usebox{\plotpoint}}
\put(447.83,300.00){\usebox{\plotpoint}}
\multiput(450,300)(20.756,0.000){0}{\usebox{\plotpoint}}
\multiput(456,300)(20.547,-2.935){0}{\usebox{\plotpoint}}
\put(468.51,299.00){\usebox{\plotpoint}}
\multiput(469,299)(20.756,0.000){0}{\usebox{\plotpoint}}
\multiput(475,299)(20.473,-3.412){0}{\usebox{\plotpoint}}
\multiput(481,298)(20.756,0.000){0}{\usebox{\plotpoint}}
\put(489.17,297.81){\usebox{\plotpoint}}
\multiput(494,297)(20.756,0.000){0}{\usebox{\plotpoint}}
\multiput(500,297)(20.473,-3.412){0}{\usebox{\plotpoint}}
\put(509.77,296.00){\usebox{\plotpoint}}
\multiput(512,296)(20.756,0.000){0}{\usebox{\plotpoint}}
\multiput(519,296)(20.473,-3.412){0}{\usebox{\plotpoint}}
\put(530.45,295.00){\usebox{\plotpoint}}
\multiput(531,295)(20.473,-3.412){0}{\usebox{\plotpoint}}
\multiput(537,294)(20.756,0.000){0}{\usebox{\plotpoint}}
\multiput(544,294)(20.473,-3.412){0}{\usebox{\plotpoint}}
\put(551.04,293.00){\usebox{\plotpoint}}
\multiput(556,293)(20.473,-3.412){0}{\usebox{\plotpoint}}
\multiput(562,292)(20.756,0.000){0}{\usebox{\plotpoint}}
\put(571.67,291.48){\usebox{\plotpoint}}
\multiput(575,291)(20.756,0.000){0}{\usebox{\plotpoint}}
\multiput(581,291)(20.473,-3.412){0}{\usebox{\plotpoint}}
\put(592.31,290.00){\usebox{\plotpoint}}
\multiput(593,290)(20.547,-2.935){0}{\usebox{\plotpoint}}
\multiput(600,289)(20.473,-3.412){0}{\usebox{\plotpoint}}
\multiput(606,288)(20.756,0.000){0}{\usebox{\plotpoint}}
\put(612.90,287.85){\usebox{\plotpoint}}
\multiput(618,287)(20.756,0.000){0}{\usebox{\plotpoint}}
\multiput(624,287)(20.547,-2.935){0}{\usebox{\plotpoint}}
\put(633.48,285.59){\usebox{\plotpoint}}
\multiput(637,285)(20.756,0.000){0}{\usebox{\plotpoint}}
\multiput(643,285)(20.473,-3.412){0}{\usebox{\plotpoint}}
\put(654.10,284.00){\usebox{\plotpoint}}
\multiput(656,284)(20.473,-3.412){0}{\usebox{\plotpoint}}
\multiput(662,283)(20.473,-3.412){0}{\usebox{\plotpoint}}
\multiput(668,282)(20.756,0.000){0}{\usebox{\plotpoint}}
\put(674.68,281.89){\usebox{\plotpoint}}
\multiput(680,281)(20.547,-2.935){0}{\usebox{\plotpoint}}
\multiput(687,280)(20.473,-3.412){0}{\usebox{\plotpoint}}
\put(695.21,279.00){\usebox{\plotpoint}}
\multiput(699,279)(20.473,-3.412){0}{\usebox{\plotpoint}}
\multiput(705,278)(20.547,-2.935){0}{\usebox{\plotpoint}}
\put(715.81,277.00){\usebox{\plotpoint}}
\multiput(718,277)(20.473,-3.412){0}{\usebox{\plotpoint}}
\multiput(724,276)(20.473,-3.412){0}{\usebox{\plotpoint}}
\multiput(730,275)(20.473,-3.412){0}{\usebox{\plotpoint}}
\put(736.32,273.95){\usebox{\plotpoint}}
\multiput(743,273)(20.756,0.000){0}{\usebox{\plotpoint}}
\multiput(749,273)(20.473,-3.412){0}{\usebox{\plotpoint}}
\put(756.90,271.68){\usebox{\plotpoint}}
\multiput(761,271)(20.547,-2.935){0}{\usebox{\plotpoint}}
\multiput(768,270)(20.473,-3.412){0}{\usebox{\plotpoint}}
\put(777.44,269.00){\usebox{\plotpoint}}
\multiput(780,269)(20.473,-3.412){0}{\usebox{\plotpoint}}
\multiput(786,268)(20.473,-3.412){0}{\usebox{\plotpoint}}
\put(797.97,266.15){\usebox{\plotpoint}}
\multiput(799,266)(20.473,-3.412){0}{\usebox{\plotpoint}}
\multiput(805,265)(20.473,-3.412){0}{\usebox{\plotpoint}}
\multiput(811,264)(20.473,-3.412){0}{\usebox{\plotpoint}}
\put(818.47,263.00){\usebox{\plotpoint}}
\multiput(824,263)(20.473,-3.412){0}{\usebox{\plotpoint}}
\multiput(830,262)(20.473,-3.412){0}{\usebox{\plotpoint}}
\put(836,261){\usebox{\plotpoint}}
\end{picture}
\end{center}
\caption{Speed of light squared from the quark dispersion relation using the
D234c and SW actions, for pseudoscalar (``$\pi_s$'') and vector
(``$\phi$'') mesons made of strange quarks. Simulations are on isotropic
lattices with $a\!\approx\!.4$\,fm and an improved gluon action. Fits shown
are to $c_0+c_4 (pa)^4$ for D234c and $c_0+c_2(pa)^2$ for SW.}
\label{fig:c2}
\end{figure}
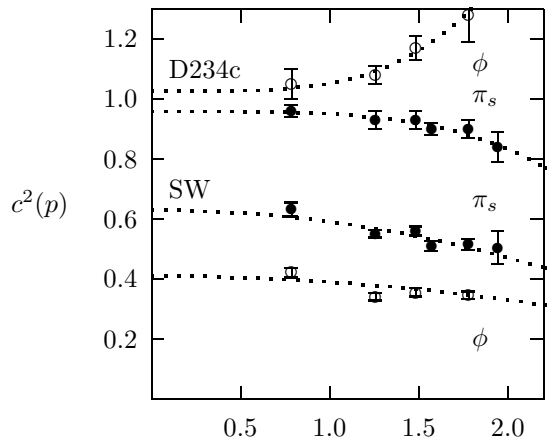

The leading correction in $M_{\rm D234c}$ beyond those in the SW formalism
is 
\be
-c_3\,\frac{a^2}{6}\sum_\mu \Delta^{(3)}_\mu\gamma_\mu.
\ee
This violates Lorentz invariance; it cancels similar violations in the
leading-order terms. Thus we tested this correction by computing
\be
c^2(\pv) \equiv \frac{E_h^2(\pv)-E_h^2(0)}{\pv^2}
\ee
where $E_h(\pv)$ is the energy of hadron~$h$ at three-momentum~$\pv$.
In a Lorentz invariant theory, $c(\pv)$ should equal~1, the speed of
light, for all~$\pv$. Results at~0.4\,fm
are shown in Fig.~\ref{fig:c2} for both the D234c and SW actions. D234c is
dramatically superior, deviating from~1 by only 3--4\% at $\pv\!=\!0$. Also
the dispersion relation is accurate to within 10\% out to three-momenta of
order~$1.5/a$. 

We could tune the coupling~$c_3$ in the
D234c action nonperturbatively until $c^2(0)\!=\!1$, thereby restoring
Lorentz symmetry through order~$a^2$.   In the tadpole-improved theory at
0.4\,fm, however, $c_3\!=\!1$  is so close to the correct value there
is no point. There is no value of $c_3$ for which both the pseudoscalar and
vector have exactly $c^2\!=\!1$; uncorrected $a^4$ errors are larger than
the radiative corrections to $c_3$ at this lattice spacing.

There are two different definitions of a hadron's mass in lattice
simulations. One is the ``static mass,'' $E_h(0)$, and the other the
``kinetic mass,'' which equals $E_h(0)/c^2(0)$. In D234c these two
definitions agree to within~3-4\%. In SW they differ by 40--60\%. The
kinetic and static masses in either formalism must be equal for zero mass
mesons, because of the axis-interchange symmetry of the actions. The
deviations seen here are because the strange quark is relatively massive at
0.4\,fm; the $\phi$~mass, for example, is~$2.1/a$ with D234c. 

Another test of the D234c action is to vary the parameter~$r$. The
$r$-dependent terms in the D234c action are approximately redundant at low
momenta, since they originate from a field transformation. We ran
simulations both at the standard value $r\!=\!1$ and at $r\!=\!2/3$ to check
this. With D234c the hadron masses changed by no more than 1\%. 
Reducing $r$ in this way lowers one of the ghost branches of the quark
dispersion relation from an energy near $2/a$ to one near~$1/a$. The
lowering of the ghost changed little in the spectrum;
our results are not strongly affected by ghost artefacts.

With SW, where the $r$~terms are less highly corrected and therefore less
redundant, we found shifts of order~1--5\% when we varied~$r$ as above.

The coefficients $c_F$, $c_3$, and $c_4$ are all renormalized by residual
quantum effects, even after tadpole improvement. We varied these couplings
to assess the importance of the potential corrections. Only $c_F$ had a
strong effect on the static hadron masses. D234c with $c_F\!=\!1$ gives
scaling errors in the $\phi$~mass of order 5\% at 0.25\,fm and 15\%
at~0.4\,fm; but a radiative correction of $0.5\alpha_V$ in $c_F$, like that
in the SW action (\eq{csw-sw}), would shift these masses by~4--8\%.
Thus to achieve precisions of order a few percent at such large lattice
spacings, we require a calculation of the order~$\alpha_V$ correction
to~$c_\sw$. This calculation will be completed in the very near future.

The highly corrected D234c action is far superior to SW for states with high
energies, either because $\pv$ is large or because $a$ is large. This makes
it potentially useful for simulations with very large lattice spacings,
large quark masses, or high-momentum states (for example, high-energy
weak interaction form factors). Anisotropic versions of D234c have also been
tested\,\cite{d234c,schladming}.

\subsection{Improved Staggered Quarks}
The standard staggered-quark action, like the SW action, has leading errors
of order~$a^2$. Some attempt has been made to correct the
$a^2$~errors\,\cite{toussaint}, but this effort included only the obvious
correction to the derivative in the action. Surprisingly, there are, in
addition, four-quark interactions that must be included at tree-level in
order~$a^2$. Tree-level improvement of a lattice action requires that we
match on-shell, tree-level, lattice scattering amplitudes with continuum QCD
at low external momentum. Normally this implies that the lattice theory has no
tree-level four-quark interactions since continuum QCD has none: in a
tree-level quark-quark scattering amplitude, low-momentum external quarks
imply a low-momentum internal gluon, and therefore these amplitudes will
match automatically if the quark-gluon vertices match. This reasoning is
incorrect for staggered quark formalisms. For example, a pair of low-momentum
quarks can scatter exchanging a {\em high-energy\/} gluon with momentum
$q=(\pi/a,0,0,0)$, creating a pair of low-momentum quarks of a different
flavor. Such flavor changing quark-quark amplitudes are absent in QCD and so
must be eliminated by adding correction terms to the action. Since the gluon
is highly virtual in a flavor-changing process, the correction term is a
four-quark operator with a coefficient of a constant times
$a^2\alpha_V(q_g)$, where the constant is computed from the scattering
amplitude and $q_g$ is the gluon momentum. 

The four-quark corrections obviously will greatly reduce the flavor-symmetry
violations observed in hadronic spectra from staggered quarks, since these
are proportional to~$a^2$. Supporting evidence also comes from
the numerical experiments described
in~\cite{toussaint}. In these experiments, 
the link operator in the standard staggered-quark
action was replaced by a smeared link. Smearing the
links inhibits the exchange of high-momentum gluons by lowering the
effective ultraviolet cutoff of the gluons; but the flavor-changing
interaction requires exchanged gluons with the highest momenta on the
lattice, $\pm \pi/a$ in one or more directions. Thus smearing should
dramatically reduce the flavor splittings; and it does\,\cite{toussaint}.

If the four-quark operators are computed and can be included in a practical
action, the staggered-quark formalism will be more accurate than the SW
action. Adding the four-quark corrections may also significantly
reduce the size of $\alpha_V$~corrections to the mass renormalization
for staggered quarks. This correction term has normal size for
$r\!\approx\!1$, but becomes quite large in the staggered-quark limit,
$r\!=\!0$. Thus the large size in this limit seems to be related to
flavor mixing. A reduction of this renormalization is important for the
determination of the strange-quark mass.

\section{Conclusions}\label{conclusions}
Techniques for simulating on coarser lattices are rapidly gaining 
acceptance within the lattice QCD community, and work at lattice spacings in
the range of 0.1--0.2\,fm is becoming widespread, again. It is likely
that some combination of perturbatively improved actions, nonperturbative
improvement, perfect actions, anisotropic lattices, Monte Carlo
Renormalization Group, and the like will allow us to simulate accurately
and efficiently on still coarser lattices in the very near future.

\end{document}